\documentclass[lettersize,journal]{IEEEtran}

\renewcommand {\marginpar}[1]{}

\newcommand {\rfig}[1]{Figure \ref{fig:#1}}

\newcommand {\rsec}[1]{Sec. \ref{sec:#1}}

\newcommand {\rsubsec}[1]{Sec. \ref{sec:#1}}

\newcommand {\beq}[1]{
                      \begin{equation}
                      \label{eq:#1} }
\newcommand {\eeq}{\end{equation}}
\newcommand {\beqno}[1]{\begin{eqnarray}
                      \nonumber}

\newcommand {\eeqno}{ && \end{eqnarray}}

\newcommand {\bear}[1]{
                       \begin{eqnarray}
                       \label{eq:#1} }

\newcommand {\bearno}[1]{
                       \begin{eqnarray}
                       \nonumber}

\newcommand {\eear}{\end{eqnarray}}
\newcommand {\eearno}{\end{eqnarray}}

\newcommand {\btab}[1]{
                       \begin{table}
                       \centering
                       \begin{tabular}{#1}}
\newcommand {\etab}[3] {
                       \end{tabular}
                       \caption[#3]{#2}
                       \label{tab:#1}
                       \end{table}
                       \vspace{.1in}}
\newcommand {\rtab}[1]{Table \ref{tab:#1}}

\newcommand {\btabular}[1]{\begin{center}
                       \begin{tabular}{#1}}
\newcommand {\etabular}{\end{tabular}
                       \end{center}}

\newcommand {\bdefin}[1]{\begin{definition}\label{def:#1}}
\newcommand {\edefin}       {\end{definition}}

\newcommand {\bpro}[1]{\begin{property}
                      \label{pro:#1} }
\newcommand {\epro}   {\end{property}}

\newcommand {\bprop}[1]{\begin{proposition}
                      \label{prop:#1} }
\newcommand {\eprop}       {\end{proposition}}

\newcommand {\blem}[1]{\begin{lemma}
                      \label{lem:#1}}
\newcommand {\elem}   {\end{lemma}}

\newcommand {\bthe}[1]{\begin{theorem}
                      \label{the:#1} }
\newcommand {\ethe}   {\end{theorem}}

\newcommand {\bcor}[1]{\begin{corollary}
                      \label{cor:#1} }
\newcommand {\ecor}   {\end{corollary}}

\newcommand{\hide}[1]{}

\newcommand {\shil}[1]{{\color{red}{#1}}}
\newcommand {\sunye}[1]{{\color{blue}{#1}}}
\newcommand {\zipeng}[1]{{\color{cyan}{#1}}}

\newcommand{\mkclean}{
   \renewcommand{\shil}[1]{}
   \renewcommand{\sunye}[1]{}
   \renewcommand{\zipeng}[1]{}
}
\mkclean

\usepackage{color}
\usepackage{amsmath,amsfonts}
\usepackage{array}
\usepackage[caption=false,font=normalsize,labelfont=sf,textfont=sf]{subfig}
\usepackage{textcomp}
\usepackage{stfloats}
\usepackage{url}
\usepackage{verbatim}
\usepackage{graphicx}
\usepackage{cite}
\usepackage{tabu}                      
\usepackage{booktabs}                  
\usepackage{lipsum}                    
\usepackage{mwe}                       
\usepackage{graphicx}
\usepackage{caption}
\usepackage{float}
\usepackage{adjustbox}
\usepackage{pifont} 
\usepackage{amsmath}
\usepackage{multirow}
\usepackage{algorithm}
\usepackage{algorithmicx}
\usepackage{algpseudocode}
\usepackage{mathptmx}                  
\usepackage{hyperref} 
\usepackage{orcidlink} 

\usepackage{titlesec}

\titlespacing{\subsection}{0pt}{*1}{*0.5}

\begin{document}

\title{GeneticPrism: Multifaceted Visualization of \\Scientific Impact Evolutions}

\author{%
  Ye Sun~\orcidlink{0000-0002-3679-483X},
  Zipeng Liu~\orcidlink{0000-0003-3567-6986},
  Yuankai Luo~\orcidlink{0000-0003-3844-7214},
  Lei Xia~\orcidlink{0009-0008-7983-9592},
  Lei Shi~\orcidlink{0000-0002-1965-2602}

\thanks{Ye Sun (sunie@buaa.edu.cn), Yuankai Luo (luoyk@buaa.edu.cn), Lei Xia (lei\_xiaaa@163.com), Lei Shi (leishi@buaa.edu.cn) are with the School of Computer Science, Beihang University, Beijing 100191, China. Lei Shi is the corresponding author.}
\thanks{Zipeng Liu (zipeng@buaa.edu.cn) is with the School of Software, Beihang University, Beijing 100191, China, and Zhongguancun Laboratory, China.}
}

\markboth{IEEE TRANSACTIONS ON VISUALIZATION AND COMPUTER GRAPHICS,~Vol.~XX, No.~XX, August~2024}%
{Shell \MakeLowercase{\textit{et al.}}: A Sample Article Using IEEEtran.cls for IEEE Journals}


\maketitle


\begin{figure*}[tb]
  \centering
  \includegraphics[width=\textwidth, clip, trim=0pt 0pt 0pt 10pt]{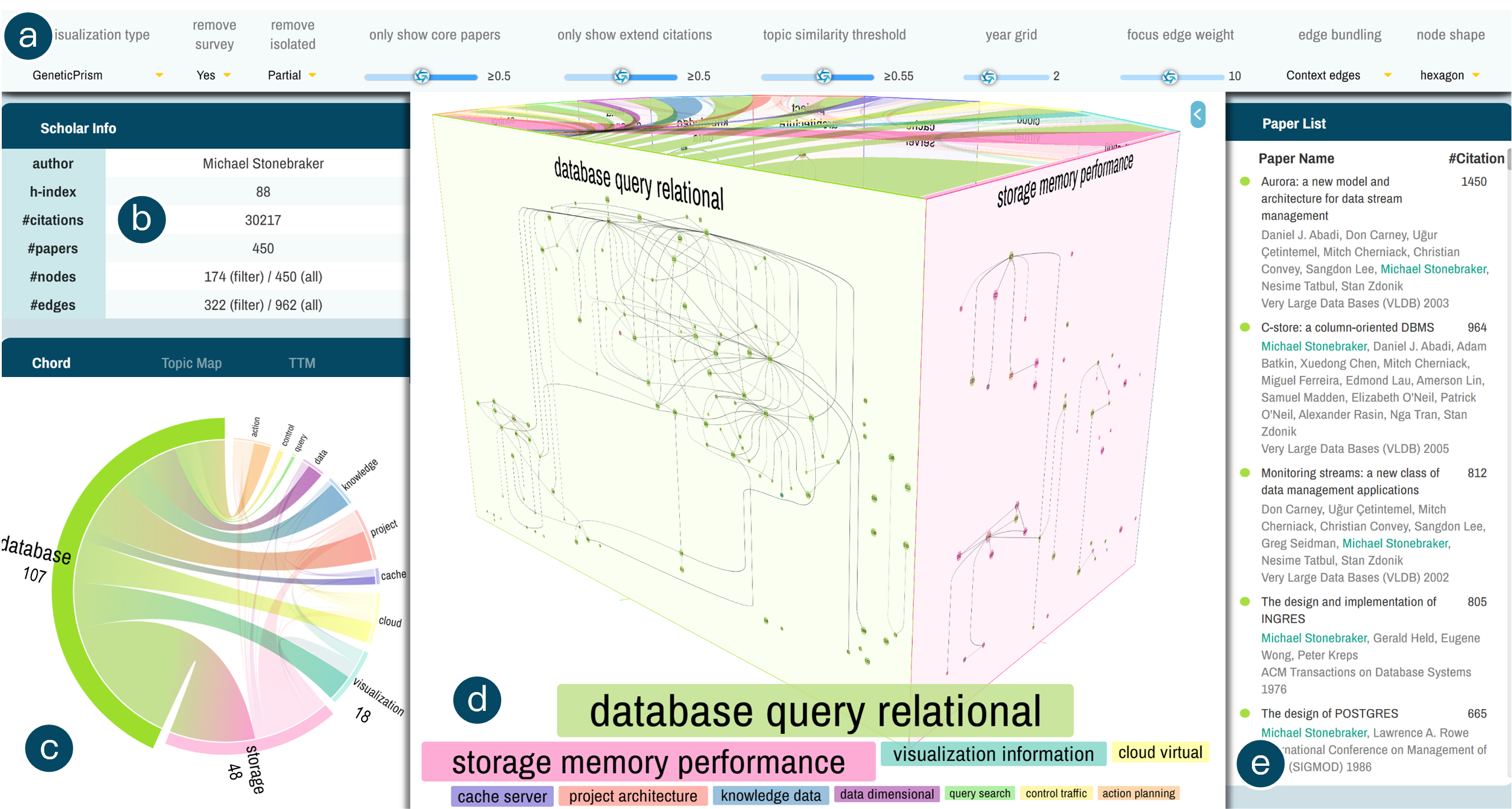}
  \vspace{-1.5em}
  \caption{GeneticPrism interface: (a) control panel; (b) scholar demographics and GeneticFlow graph statistics; (c) chord diagram to display inter-topic citation influence interactions; (d) main GeneticPrism (current) or GeneticScroll designs visualizing Shirato graphs; (e) paper list and detailed info panel.
  }
  \label{fig:GeneticPrismSystem}
  \vspace{-1em}
\end{figure*}

\begin{abstract}
Understanding the evolution of scholarly impact is essential for many real-life decision-making processes in academia, such as research planning, frontier exploration, and award selection. Popular platforms like Google Scholar and Web of Science rely on numerical indicators that are too abstract to convey the context and content of scientific impact, while most existing visualization approaches on mapping science do not consider the presentation of individual scholars' impact evolution using curated self-citation data. This paper builds on our previous work and proposes an integrated pipeline to visualize a scholar's impact evolution from multiple topic facets. A novel 3D prism-shaped visual metaphor is introduced as the overview of a scholar's impact, whilst their scientific evolution on each topic is displayed in a more structured manner. Additional designs by topic chord diagram, streamgraph visualization, and inter-topic flow map, optimized by an elaborate layout algorithm, assist in perceiving the scholar's scientific evolution across topics. A new six-degree-impact glyph metaphor highlights key interdisciplinary works driving the evolution. The proposed visualization methods are evaluated through case studies analyzing the careers of prestigious Turing award laureates and a major visualization venue. \sunye{An online visualization system will be released upon acceptance.}
\end{abstract}


\section{Introduction}

\maketitle


Understanding and analyzing the evolution of scientific impact is crucial for comprehending the intricate nature of scholarly contribution and influence. Many real-world tasks such as award selection, research topic planning \cite{niksirat2023effective} and frontier exploration \cite{fortunato2018science}, as well as tenure evaluation \cite{schimanski2018evaluation} could benefit from such study.\shil{Could use more latest citations, should also be highly cited ones.} Visualization is indispensable in delineating the evolution of scientific topics at the scholar level \cite{xiao2023geneticflow}\cite{wang2018visualizing}, institutional scale \cite{heimerl2015citerivers}\cite{van2014visualizing}, and for entire research fields \cite{nature_150_interactive}.\shil{Need to add these citations, use newer and impactful citations.} Compared to abstract indicators like h-index and impact factor popular in scholarly platforms (Google Scholar \cite{google_scholar}, Web of Science \cite{WoS}, etc.), the visualization of collaboration networks \cite{wu2015egoslider}, co-citation/bibliographic graphs \cite{huang2019eiffel}, and topic content evolutions \cite{heimerl2015citerivers} analyze contextual data better, uncover hidden scientific patterns, and provide detailed reasoning.

In the literature, mapping the evolution of sciences has been the research community's goal ever since Eugene Garfield and others established the famous Institute for Scientific Information decades ago based on citation data \cite{garfield2009science}. Later on, CiteSpace (II) pioneered the visualization of research fronts and intellectual base using the concept of co-citations \cite{chen2016citespace}\cite{chen2006citespace}. Nature also launched a visualization project recently to mark the journal's 150th anniversary, illustrating 19 million articles on their co-citation networks \cite{nature_150_interactive}. Till now, numerous systems and software have been proposed for mapping science, using a variety of information visualization techniques such as networks \cite{boyack2005mapping}, maps \cite{fried2014maps}, matrices \cite{wang2018visualizing}, etc. Our work builds over the latest GeneticFlow design by Xiao and Shi \cite{xiao2023geneticflow}, who utilize curated self-citation data to illustrate the scientific impact evolution of individual scholars, a task that very few existing visualization approaches consider.

The GeneticFlow platform has been extended to support the visual profiling of more than 100k scholars from 11 research areas of computer science \cite{GeneticFlow}. The technique works well on most scholars with a small to medium number of publications in their entire career. However, for top scholars accumulating hundreds of papers and thousands of self-citation links, GeneticFlow adopting classical visualization algorithms suffers from huge visual clutter and edge crossings, a known issue for large graph visualization. Moreover, the original design improvises the display of the most crucial topic information with a plain color-coding on paper nodes. Both intra-topic and inter-topic impact evolutions are hardly perceived due to the complex nature of topic distribution over a scholar's career.

In this work, we are motivated by the observation of a universal pattern in scholars' scientific evolution: their works on the same research topic are normally more interconnected (by citation links) than those on different topics. A straightforward proposal would be splitting a top scholar's GeneticFlow graph into multiple topic-based facets and displaying them separately. Visualizing each topic facet with a much smaller number of papers and citations in a standalone view will be more effective and intuitive for analysis. This simple approach, though compelling, still faces multiple technical challenges. First, how can we visually accommodate a scholar's potentially 10+ topic facets in a traditional 2D display space? Second, while inter-topic citations are more infrequent, they are at least as crucial as intra-topic citations because the key information of topic evolution is encapsulated there. How do we visualize a scholar's scientific evolution in a topic-based manner while still uncovering topic-level interactions? Third, the research topic development of a scholar is not linear in that many works overlap accross multiple topics. How do we display this critical evidence of a scholar's scientific evolution? We make the following contribution to tackle the challenges above.
\begin{itemize}
    \item We introduce GeneticPrism (\rfig{GeneticPrismSystem}(d)), a novel 3D prism-shaped visual metaphor, to integrate multiple topic facets of a scholar in the same display. Complemental designs, including a topic chord diagram (\rfig{GeneticPrismSystem}(c)), customized animations and interactions, are also presented to illuminate the overview of a scholar's scientific impact evolution. We also propose an end-to-end data analysis and topic modeling pipeline to support GeneticPrism visualization (\rfig{Pipeline});
    \item On mapping scholar's scientific contribution from a single topic facet, we introduce the GeneticScroll design, which orchestrates a hierarchical graph visualization layer for intra-topic impact evolution with a flow map and streamgraph visualization layer on inter-topic citation influences. An elaborate graph layout algorithm is proposed to coordinate the two layers and minimize edge crossings and visual clutter. We also propose a new six-degree-impact glyph design highlighting interdisciplinary works on GeneticScroll. Note that throughout this work, by interdisciplinary, we mean overlaps among fine-grained research topics, not using its original definition among high-level research areas;
    \item We implement the GeneticPrism system on Microsoft Academic Graph (MAG), the largest open academic database. Case studies on two prestigious Turing award laureates demonstrate the effectiveness of GeneticPrism by clearly delineating their major research threads from multiple topic facets, visually reasoning on important topic evolution patterns, and identifying key interdisciplinary papers that drive their impact evolution. It is also shown that the approach can be extended to support impact evolution analysis at the academic venue level.
\end{itemize}

\section{Related Work}

\subsection{Visualizing Scientific Impact Evolution}

Visualizing scientific impact evolution involves understanding and presenting a scholar's contributions over time and examining how topics, authors, and scientific trends develop and impact their fields. A list of related works and comparisons is shown in Table~\ref{tab:visualizing_impact}. Topic-based visualization is crucial as it allows for identifying and tracking thematic developments. The related works can be classified into three categories: visual factor analysis for career development, ego-centric scholar visualization, and citation pattern visualization.

\begin{table}[tb]
  \caption{%
    A comparative analysis of related works on visualizing scientific impact evolution. The final row gives GeneticPrism, our proposed method. \ding{51}: Yes, $\circ$: Partial.%
  }
  \label{tab:visualizing_impact}
  \scriptsize%
  \centering%
  \renewcommand{\arraystretch}{1.2} 
  \begin{tabu}{%
      X[1.5,l]%
      X[0.4,c]%
      X[0.4,c]%
      X[0.4,c]%
      X[0.4,c]%
      X[0.4,c]%
      X[2.7,l]%
      X[0.5,c]%
    }
    \toprule
    \rotatebox{45}{Method} & \rotatebox{45}{Topic-based} & \rotatebox{45}{Temporal} & \rotatebox{45}{Career analysis} & \rotatebox{45}{Domain insight} & \rotatebox{45}{Impact evolution} & \rotatebox{45}{Visual Technique} & \rotatebox{45}{Year} \\ \midrule
    GeneticFlow~\cite{xiao2023geneticflow}          & \ding{51}            & \ding{51}        & \ding{51}                 &            & \ding{51} & node-link                & 2023 \\
    PubExplorer~\cite{yu2023pubexplorer}          & $\circ$          & \ding{51}        & $\circ$                 &       \ding{51}      & $\circ$ & state map, histogram                & 2023 \\
    ACSeeker~\cite{wang2021seek}             & $\circ$           & \ding{51}        & \ding{51}                 &            &           & timeline, matrix                       & 2022 \\
    Eiffel~\cite{huang2019eiffel}               & $\circ$          & \ding{51}        &                           & \ding{51}  &     $\circ$     & Flow map                       & 2020 \\
    Nature150~\cite{nature_150_interactive}               & $\circ$          &  $\circ$        &                           & $\circ$  &           & 3D node-link                       & 2019 \\
    ImpactVis~\cite{wang2018visualizing}            & $\circ$           & \ding{51}        &               & \ding{51}  &  $\circ$        & matrix, glyph, stacked bar chart                         & 2018 \\
    Vis Author Profiles~\cite{latif2018vis}  &    $\circ$     & \ding{51}        & \ding{51}                 &            &           & text, stacked bar chart            & 2018 \\
    CiteRivers~\cite{heimerl2015citerivers}           & \ding{51}           & \ding{51}        &                           & \ding{51}  &   $\circ$       & streamgraph                    & 2016 \\
    ScholarPlot~\cite{majeti2020scholar}         &                     & \ding{51}        & \ding{51}                 &            &           & scatter plot                   & 2016 \\
    
    VEGAS~\cite{shi2015vegas}                & $\circ$            &                  &                           & \ding{51}  & $\circ$  & node-link                    & 2015 \\
    VOSviewer~\cite{van2014visualizing}             & $\circ$           &        & $\circ$                 &    \ding{51}        &    $\circ$       & node-link                       & 2014 \\
    Pathway~\cite{wu2013visual}         &                     & \ding{51}        & \ding{51}                 &            &           & area chart                   & 2013 \\
    \midrule
    \textbf{GeneticPrism}  & \textbf{\ding{51}}  & \textbf{\ding{51}} & \textbf{\ding{51}}       & \textbf{\ding{51}} & \textbf{\ding{51}} & \textbf{3D prism, node-link, chord, streamgraph, flow map} & \textbf{Ours} \\ \bottomrule
  \end{tabu}%
  \vspace{-2em}
\end{table}

Visual factor analysis includes works that analyze and visualize multiple factors influencing academic careers. ACSeeker~\cite{wang2021seek} explores potential factors of academic success across different career stages. ImpactVis~\cite{wang2018visualizing} uses a matrix form to visualize individual careers and domain trends. Vis Author Profiles~\cite{latif2018vis} combines text and visualization for insights into individual and domain impacts. ScholarPlot~\cite{majeti2020scholar} and Pathway~\cite{wu2013visual} use scatter plots and area charts to depict career trajectories. These works lack detailed topic-based or domain-specific analysis.

Ego-centric scholar visualization focuses on a single scholar's academic impact and collaboration network. EgoSlider~\cite{wu2015egoslider} uses a timeline-based visualization for evolutionary collaboration networks. EgoLines~\cite{zhao2016egocentric} employs a ``subway map'' metaphor for temporal dynamics. Reitz~\cite{reitz2010framework} uses a scholar's ego-centric node-link diagram for temporal distributions of joint works. MENA~\cite{he2016mena} represents ego-centric networks as dynamic graphs in small multiples. Fung et al.~\cite{fung2016design} suggest a botanically inspired tree visualization summarizing collaborations over time. Episogram~\cite{cao2015episogram} presents a novel visualization tool for summarizing egocentric social interactions, enabling users to explore and compare social behaviors over time. This class of works focuses on the collaboration of scholars but not on broader impact evolutions.

Citation pattern visualization highlights citation patterns and the influence of individual scholars. Eiffel~\cite{huang2019eiffel} applies triple summarizations and uses a dynamic flow map for hierarchical and temporal changes of a paper's citation influence. CiteRivers~\cite{heimerl2015citerivers} employs a streamgraph for temporal trends in domain-specific impacts. VEGAS~\cite{shi2015vegas} employs visual influence graph summarization to enable the exploration of highly influential papers. The Nature journal~\cite{nature_150_interactive} offers galaxy and light cone views for co-citation networks. VOSviewer~\cite{van2014visualizing} uses density and network visualizations for citation analysis. PubExplorer~\cite{yu2023pubexplorer} employs citation pattern analysis and science maps to explore research trends and collaborations. These methods highlight citation patterns but may lack detailed temporal and topic-based insights.

GeneticFlow (GF)~\cite{luo2023impact, xiao2023geneticflow} introduces methods for constructing graph-based scholar profiles, illustrating academic trajectories and scholarly impact. This structured representation of a scholar's scientific development emphasizes the evolution of research topics. Still, it does not work well for top scholars with abundant publications and citation influence. Additionally, its intrusive topic modeling, which assigns only one topic per paper, reduces the possibility of understanding domain insight in complex, multi-topic scenarios.

Despite their contributions, current approaches to visualizing scientific impact evolution face scalability issues and challenges in visualizing overlapping topics. In comparison, GeneticPrism distinguishes itself by integrating topic-based, temporal, individual career, and domain-specific visualizations using a combination of node-link graphs, chord diagrams, streamgraphs, and flow maps, providing a holistic solution of scientific impact visualization across various dimensions.


\subsection{Visualizing Multilayer Networks}

Multilayer network visualization represents systems where entities interact across multiple contexts or dimensions~\cite{de2022multilayer}. Nodes can belong to multiple layers, with edges existing within or spanning across layers. This kind of method benefits fields like biology, sociology, and digital humanities. Recent surveys~\cite{mcgee2019state, filipov2023we} highlight the need for advanced visualization techniques to handle the complexity and volume of multilayer network data.

Several innovative approaches have been developed. MuxViz~\cite{de2015muxviz} provides a platform for analyzing and visualizing multilayer networks using techniques like adjacency matrices and node-link diagrams. Detangler~\cite{renoust2015detangler} offers coordinated views to explore inter-layer relationships. Multilayer Graph Edge Bundling~\cite{bourqui2016multilayer} addresses edge clutter by grouping edges with common paths. Refinery~\cite{kairam2015refinery} supports cross-layer connectivity tasks through associative browsing. HybridVis~\cite{liu2017hybridvis} integrates multiple visualization techniques for large multivariate networks, providing a flexible approach to visualize different data aspects. NetworkAnalyst~\cite{xia2015networkanalyst} uses chord diagrams for inter-layer relationships in gene expression data. Dynamic Communities~\cite{vehlow2015visualizing} visualizes the evolution of community structures within multilayer networks. g-Miner~\cite{chen2018g} offers interactive techniques for exploring and manipulating graph data, including creating and refining groupings within layers. Shirato et al.~\cite{shirato2023exploring} enhance the understanding of complex temporal relationships in multivariate time series through advanced visualization techniques across multiple dimensions. GraphDice~\cite{bezerianos2010graphdice} extends ScatterDice~\cite{elmqvist2008rolling} for multivariate networks, providing a consistent interface for visualizing node and edge attributes. Similar to GeneticPrism, GraphDice supports exploring complex multilayer networks through filtering and partitioning based on various attributes but focuses more on interactive exploration of network dimensions and does not present the influence among groups.

The survey by McGee et al.~\cite{mcgee2019state} categorizes tasks into cross-layer connectivity, cross-layer entity comparison, layer manipulation, and layer comparison. However, tasks addressing inter-relationships between layers, such as cross-layer flows, are rarely explored. GeneticPrism addresses this gap by visualizing inter-layer relationships through a suite of novel visualization designs. Furthermore, GeneticPrism supports overlapping topic visualization that simultaneously displays nodes appearing in multiple network layers.

\section{Scientific Impact Evolution}

\begin{figure*}[tb]
  \centering
  \includegraphics[width=\textwidth, clip, trim=10pt 10pt 10pt 40pt]{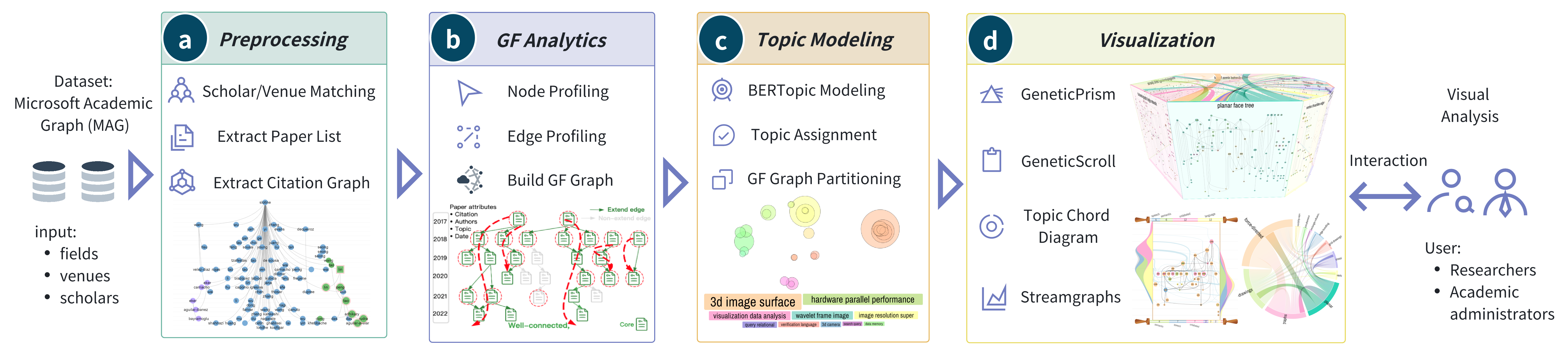}
  \vspace{-1.5em}
  \caption{GeneticPrism pipeline over MAG to illustrate scientific impact evolution: (a) data pre-processing; (b) GF analytics to build citation influence graph; (c) topic modeling for multifaceted analysis; (d) visualization designs.\shil{Should annotate subfigure abcd, scientific officers => academic administrators.}\sunye{Revised}
  }
  \label{fig:Pipeline}
  \vspace{-1em}
\end{figure*}

The proposed pipeline to illustrate scientific impact evolution is shown in \rfig{Pipeline}. The raw academic data after preprocessing is first modeled by GF analytics into citation influence graphs, or GF graphs in short. A new topic modeling stage is introduced to assemble multiple topic facets on each GF graph. GeneticPrism and other visual metaphors are then designed for the multi-view visualization.

\subsection{Academic Data Source and Pre-processing}

\begin{table}[t]
\centering
  \caption{
     GF graph statistics used in this work, including award recipients, top scholars in CS sub-fields, and academic venues.
  }
  \label{tab:graph_statistics}
  \scriptsize
    \begin{tabu} to 0.5\textwidth {%
          X[1.4,l]%
          X[2.4,l]%
          X[0.6,c]%
          X[0.5,c]%
          X[0.6,c]%
          X[0.6,c]%
        }
    \toprule
    \textbf{Category} & \textbf{Name} & \textbf{\#Authors} & \textbf{\#Topic} & \textbf{\#Papers} & \textbf{\#Links} \\
    \midrule
    \multirow{2}{*}{\parbox{1.5cm}{Award\\Recipients}} & ACM Fellows & 1306 & 119 & 259526 & 571419 \\
    & Turing Award Winners & 76 & 119 & 15575 & 30614 \\
    \midrule
    \multirow{2}{*}{\parbox{2cm}{Top Scholars in\\11 CS Sub-fields}}
    & Graphics \& Visualization & 6109 & 81 & 113597 & 130650 \\
    & $\cdots\cdots$ & $\cdots$ & $\cdots$ & $\cdots$ & $\cdots$ \\
    \midrule
    \multirow{1}{*}{\parbox{2cm}{Academic\\Venues}} & {International Symposium on Graph Drawing} & N/A & 28 & 1703 & 3128 \\
    \bottomrule
\end{tabu}
\vspace{-1em}
\end{table}

Our primary data source is MAG \cite{microsoft_academic_graph}, the largest open academic database nowadays, with over 237 million papers, 240 million authors, and 1.63 billion citations from all research disciplines till Sept. 2021. As shown in \rtab{graph_statistics}, the raw MAG data is pre-processed to obtain the citation graph of academic entities in three categories: award recipients, including Turing award winners and ACM fellows, top scholars in 11 CS sub-fields, and an academic venue of GD.

For award recipients, their scholar lists are collected from the Internet; for top scholars in a sub-field, their scholar lists are ranked by h-index defined in MAG over all papers in each sub-field (e.g., the 6109 scholars in graphics\&visualization with a h-index above 5). Scholar name matching and disambiguation are conducted to ensure that the correct recipient/scholar is detected from our dataset and that all their papers in MAG are extracted. For the GD venue, we simply discover its venue ID in MAG and retrieve all the papers linked to that ID. After obtaining the paper list for each recipient/scholar/venue, we fetch all citation links among each list of papers.

\subsection{GeneticFlow Graph Model}

As shown in \rfig{Pipeline}(b), the pre-processed academic data is modeled by GF analytics \cite{luo2023impact} to compute a GF graph for each recipient/scholar/venue. 
The GF graph is designed to represent the evolution of a scholar/venue's scientific impact and contribution. Take the GF analytics of a scholar \( s \) as an example. His/her full GF graph is defined by \( G = \{V, E\} \), where \( V \) denotes the set of \( n \) paper nodes authored by \( s \) and \( E \) denotes the set of \( m \) reversed self-citation edges among \( V \) indicating citation influence links. Each node \( v_i \) is further associated with a timestamp \( t \) (publication year by default), an ordered list of paper co-authors \( A = \{a_1, a_2, \ldots\} \), and an extra set of paper attributes \( \Phi \). Each edge \( e_{ij} = (v_i, v_j) \) is derived from a citation from paper \( v_j \) to \( v_i \).

To represent the main component of a scholar's citation impact evolution, the concept of core GF graph is proposed, which is a subgraph \( G^* \) of \( G \) that best represents the impact of scholar \( s \). The core GF graph profiling problem is decomposed into two sub-problems: node profiling and edge profiling.

\textbf{Node profiling} involves detecting the set of core papers \( V^* \subseteq V \) published by the scholar \( s \). The exact algorithm is based on two assumptions: a paper's contribution is unequally credited to all authors by author order unless the paper is alphabetically ordered, and an author's contribution to the paper is also credited to their advisor if only: (a) the advisor is a co-author of the paper, and (b) the advisor-advisee relationship is active at the publication date of the paper.

With these assumptions, the contribution that the \( k \)-th author \( a_k \) makes on a paper can be quantified by:
\begin{equation}
p_{\text{cont}}(a_k) = \max \left(\frac{1}{k}, \max_{\forall l \neq k} \frac{p_{AA}(a_k, a_l, t)}{l}\right)
\label{eq:author}
\end{equation}

\noindent Here the popular harmonic credit allocation scheme \cite{hagen2008harmonic} is adopted
in that the \(k\)th author takes credit of \( 1/k \). \( p_{adr}(a_k, a_l, t) \) denotes the probability of \( a_k \) being the advisor of  \( a_l \) at time \( t \).

To detect the advisor-advisee relationship, an unsupervised, human-interpretable algorithm is introduced, which estimates the advisor-advisee probability between \( a_k \) and \( a_l \) by:
\begin{equation}
\begin{aligned}
p_{adr}(a_k, a_l, t) &= \frac{N_{a_k}(0, t) - N_{a_k, a_l}(0, t)}{N_{a_k, a_l}(0, t)} \\
p_{ade}(a_k, a_l, t) &= \max_{\substack{t_0 \leq t \leq t_1 \\ t_1 - t_0 \geq S_0 \\ \text{numerator} \geq S_{\text{adr}}}} \frac{\sum_{t_0 \leq t \leq t_1} \hat{N}_{a_k, a_l}(t)}{\hat{N}_{a_l}(t_0, t_1)} \\
p_{AA}(a_k, a_l, t) &= \min(1.0, p_{adr}(a_k, a_l, t)) \times \min(1.0, p_{ade}(a_k, a_l, t))
\end{aligned}
\label{eq:relationship}
\end{equation}

\noindent where \( N_{a_k}(0, t) \) is the number of papers published by \( a_k \) until time \( t \), and \( N_{a_k, a_l}(0, t) \) is the number of papers co-authored by \( a_k \) and \( a_l \) until time \( t \), \( \hat{N}_{a_k, a_l}(t) \) is the number of major papers of a co-authored by \( a_k \) and \( a_l \) at time \( t \), and \( \hat{N}_{a_l}(t_0, t_1) \) is the number of major papers by \( a_l \) in the period \([t_0, t_1]\). By default, a major paper of \( a_l \) means he/she is the top-3 author.




\textbf{Edge profiling} involves detecting the set of core citation edges \( E^* \subseteq E \) that represent the evolution of the scholar's scientific contribution. The approach focuses on classifying self-citation links according to an established taxonomy and then extracting the class of extend-type citations. To achieve this goal, a supervised learning algorithm is applied to a labeled dataset to classify citation types. Twenty features are selected and input to an Extra-Tree model. The model achieves an F1 score of 0.646 with 10-fold cross-validation. The features used in classification include metadata of cited and citing papers, citation network features, temporal correlation measures, and lexical patterns extracted from the citation context and full text. More details can be found in the original paper \cite{luo2023impact}.




\subsection{Topic Modeling}
\label{sec:Topic}

To resolve the scalability issue in visualizing very large GF graphs, we propose to slice a scholar's GF graph into multiple sub-graphs using topic modeling methods. It is observed that the citation influence graph inside each topic is usually much smaller yet more interpretable, as shown by the case studies in \rsec{Case}. The missing interactions among\sunye{Removed: topic-based} GF sub-graphs can be displayed by elaborate visualization designs in \rsec{Vis}.


The latest neural topic modeling method BERTopic \cite{grootendorst2022bertopic} is applied. The BERTopic pipeline includes BERT embedding to represent each document with a dense, high-dimensional vector, UMAP projection \cite{mcinnes2018umap} to map all document vectors within the same low-dimensional space, and then HDBSCAN clustering \cite{rahman2016hdbscan} to detect topics from documents. A list of keywords and an embedding vector is computed for each detected topic. As a single scholar's GF graph contains limited papers, resulting in sparse topics, we apply BERTopic to the set of papers corresponding to one row in \rtab{graph_statistics}. For each paper, the title, abstract, and index words are aggregated into one document for topic modeling. Both unigram and bigram schemes are allowed in learning topic keywords.


The default BERTopic implementation does not support overlapping topics because each paper is assigned to only one topic cluster. To identify interdisciplinary research, we introduce a customized topic assignment scheme so that one paper can belong to multiple topics. In more detail, the topic assignment is based on the cosine similarity between each paper's embedding vector and the detected topic embedding vector. By imposing a lower-bound similarity threshold adjustable in the visualization interface, each paper is assigned to all the topics with a similarity above that threshold. The topic assignment helps slice one GF graph into multiple topic-based sub-graphs in that all papers belonging to a topic and their citation links in between will form a GF sub-graph for the topic. Note that the number of paper nodes in all topic sub-graphs of a scholar will be larger than that in the original GF graph due to topic overlapping.

\subsection{User and Task Characterization}

Building over faceted GF graphs, GeneticPrism visualizations are designed to effectively delineate the scientific impact evolution of key scholars and academic venues. Our technique targets two types of users:
\begin{itemize}
\item \textbf{Researchers} who aim to analyze the academic development of themselves and other scholars from a data-driven perspective. The lessons learned and patterns detected can potentially assist them in future topic planning, collaboration building, and research frontier understanding;
\item \textbf{Academic administrators} with regular responsibility for reviewer recruitment, tenure evaluation, award/project selection, or even agenda-setting for a research field. The quantitative approach by GeneticPrism serves as a nice complement to the traditional way by subjective peer reviews.
\end{itemize}

To fulfill the requirements of these users, GeneticPrism supports the following tasks:
\begin{itemize}
\item \textbf{T1: temporal and topical overview of a scholar's scientific impact.} On the first hand, most users will need an overall understanding of a scholar's research impact, including the research contribution across topics and over time. The design should allow a macroscopic view with a time dimension that can juxtapose and compare the scholar's impact evolution on multiple topics;
\item \textbf{T2: detailed analysis of a scholar's impact evolution on a single research topic.} For any specific topic, e.g., the one that the user is currently working on or the administrator is overseeing, s/he will need to drill down to details to complete their job. The low-level tasks include but are not limited to identifying key papers, understanding research threads/clusters, and predicting topical/statistical trends;
\item \textbf{T3: discover influence and interaction patterns among multiple research topics of a scholar.} For topic planning and research outlook, users will need to figure out the causal relationship among multiple research topics of a scholar. This can be achieved through the visual analysis of citation influence patterns among papers belonging to these topics;
\item \textbf{T4: identifying key interdisciplinary papers and their influence patterns.} During topic transitions of a scholar, there usually is not a clear-cut pattern between old and new topics. It is crucial for our users to identify the key papers that conduct interdisciplinary research on these transitional topics. The context of this research can help explain the reasoning behind the process of scientific evolution.
\end{itemize}

\section{Visualization}
\label{sec:Vis}

\begin{figure*}[tb]
  \centering
  \includegraphics[width=\textwidth, clip, trim=10pt 300pt 10pt 100pt]{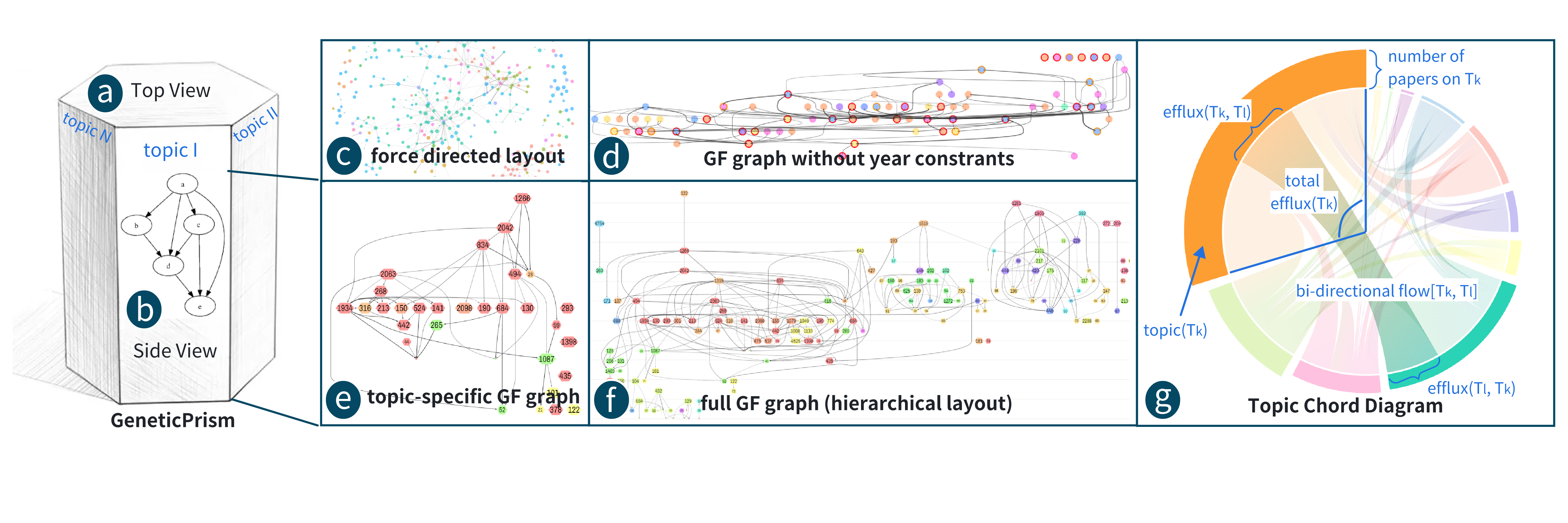}
  \vspace{-1.5em}
  \caption{%
  	GeneticPrism design and its alternatives: (a) top view by a polygon chord diagram design; (b) side view visualizing the GF sub-graph on a topic; (c)(d)(e) alternative designs; (g) chord diagram to display inter-topic interactions.\shil{Laterial =\> Vertical, remove (Base). Should tag ``topic I'', ``topic II'', etc., in the side view of (b)}\sunye{fixed}\shil{figure(g)'s annotations should be updated according to the new text in the paragraphs below.}\sunye{fixed}
  }
  \label{fig:GeneticPrism}
  \vspace{-1em}
\end{figure*}

Over GF analytics and topic modeling to illustrate the evolution of the scientific impact of key scholars and academic venues. As shown in \rfig{GeneticPrismSystem}, the interface is composed of five panels: a visualization controller to config the entire system (\rfig{GeneticPrismSystem}(a)), a demographics panel to display statistics of the current scholar (\rfig{GeneticPrismSystem}(b)), a topic interaction panel to present high-level citation influence patterns among topics (\rfig{GeneticPrismSystem}(c)), a paper list panel to detail high-impact papers of that scholar (\rfig{GeneticPrismSystem}(e)), and finally the main panel in the middle to visualize GF graphs. In the default overview mode, the GeneticPrism visualization is displayed (\rfig{GeneticPrismSystem}(d)); in the other per-topic mode, the main panel will switch to the GeneticScroll visualization (\rfig{GeneticScroll}(a)).


\subsection{GeneticPrism}

In a default mode, the main visualization panel of our system features the GeneticPrism design that fulfills the T1 overview task defined above. As shown in \rfig{GeneticPrism}(a)(b), a 3D prism-shaped metaphor is adopted, which surfaces two types of views for visualization: multiple vertical faces (i.e., side views) to display the GF sub-graph on each topic for T2, and the horizontal face in the top (i.e., top view) to display the inter-topic citation influence graphs for T3. The entire visualization is drawn in semi-transparent color so that several topic-based GF sub-graphs can be exhibited simultaneously for overview and comparison purposes. From another perspective, the GeneticPrism metaphor is designed as a mimic of a cultural monument that symbolizes the lasting and impactful scientific contribution of prestigious scholars.

On the prism metaphor, the width of each side view represents the total number of papers in the corresponding topic. \sunye{Removed: outbound citation influence links (i.e., the efflux) of the corresponding topic. The chord diagram represents, but prism does not}\shil{should double check this claim.} By this design, the top view becomes an irregular polygon inappropriate for presenting the full inter-topic interactions. Also, because the top face is normally perceived in a sloped view with a narrow-angle, we only depict a snapshot of inter-topic interactions there and visualize the full topic interaction graph in another panel of the main interface, using the chord diagram design described in Sec. \ref{sec:Vis} Part B (\rfig{GeneticPrism}(g)).

In more detail, the side view of GeneticPrism visualizes the per-topic GF sub-graph using a hierarchical layout approach that maps the paper publication year to the vertical axis. This elaborate design can reveal the scientific impact evolution of a scholar from a topic-based perspective. The popular GraphViz library \cite{ellson2002graphviz} implementing Sugiyama-style hierarchical layout algorithm is adopted. To customize the algorithm in our scenario, we bypass its first layer assignment step since each paper node has been fixed to the layer corresponding to its publication year. The following node reordering and coordinate computation steps remain intact from that of the original algorithm. After node layout, the citation influence links among papers are drawn in a third-order Bézier curve for aesthetic visualization, using additional dummy nodes to ensure smoothness.\shil{Check how to spell Bezier in latex.}\sunye{checked}

The GeneticPrism design is also equipped with multimodal interactions. Initially, the GF graph data can be queried by multiple filters (\rfig{GeneticPrismSystem}(a)) to control the number of nodes and edges, as well as the paper-topic affinity threshold. The 3D prism supports zoom\&pan and can be rotated upon mouse drag to focus on topics more interesting to the current user. Another animation mode is also supported in that the prism will rotate around its center at a constant rate, presenting a dynamic overview for better analysis. Finally, double-clicking on a side view will drill down to the GeneticScroll visualization of the corresponding topic.

\textbf{Alternative design} of GeneticPrism's side view includes:
\begin{itemize}
    \item \emph{Force-directed layout (\rfig{GeneticPrism}(c))}: the non-hierarchical graph layout misses the opportunity to illustrate the scientific evolutions of a scholar, which are in most cases hierarchically and temporally developed.
    \item \emph{GF graph without year constraints (\rfig{GeneticPrism}(d))}: when the hierarchical layout does not use publication year as the vertical axis, the resulting visualization has a height equaling the diameter of the directed graph. For typical GF graphs, citation influence chains rarely exceed five links, so the display will suffer from a considerable aspect ratio, downgrading the utility of screen size.
    \item \emph{Full GF graph (\rfig{GeneticPrism}(f))}: the original GF graph visualization in the GeneticFlow interface displays paper nodes from all topics in the same view. As mentioned, this classical design suffers from the unsolved issue of large complex graph visualization.
\end{itemize}

\subsection{Topic Chord Diagram}


To satisfy the T3 task of analyzing inter-topic interactions, a chord diagram design is introduced and presented as a standalone view in the interface (\rfig{GeneticPrismSystem}(c)). The design illustrates the citation influence graph among all derived research topics of a scholar. Denote these topics as $T_1, T_2, \ldots, T_N$, each corresponding to a topic node. The directed flow from topic $T_k$ to topic $T_l$ denotes the efflux citation influence from \(T_k\) to \(T_l\), with the flow rate computed as the total number of citation influence links from \(T_k\) to \(T_l\):
\begin{equation}
\begin{aligned}
efflux(T_k, T_l) = | \{ e_{ij} \mid e_{ij} \in E, &v_i \in V(T_k) \wedge v_i \notin V(T_l) \wedge \\
&v_j \in V(T_l) \} |
\end{aligned}
\label{eq:efflux}
\end{equation}
where $|\cdot|$ denotes the size of a set, \( G = \{V, E\} \) denotes the GF graph, $e_{ij}$ denotes the citation influence link from paper $v_i$ to $v_j$ reversed from the original citation link of $(v_j, v_i)$. It is required that the source paper node $v_i$ only belong to the source topic $T_k$, not the destination topic $T_l$, a definition consistent with the design of GeneticScroll visualization.

On the chord diagram design, as shown in \rfig{GeneticPrism}(g), each arc in the circumference represents a topic, and each chord between two arcs represents the bi-directional flow between two topics. Each arc is color-coded according to the ColorBrewer's suggestion for qualitative scales \cite{brewer2003colorbrewer}, and the chord applies transition color from the source topic to the destination topic. The thickness of an arc visually encodes the total number of papers on that topic. The arc’s angle in the circular layout is proportional to the total efflux from the current topic to all other topics of the scholar by Equation \ref{eq:efflux}. The arc is further split into endpoints for all connecting chords to destination topics, with the circular angle of each endpoint assigned by the efflux to each other topic. Again, the width of a chord smoothly transits between the endpoints of the two connecting arcs.

\subsection{GeneticScroll}

\begin{figure}[tb]
  \centering
  \includegraphics[width=\columnwidth, clip, trim=10pt 12pt 10pt 10pt]{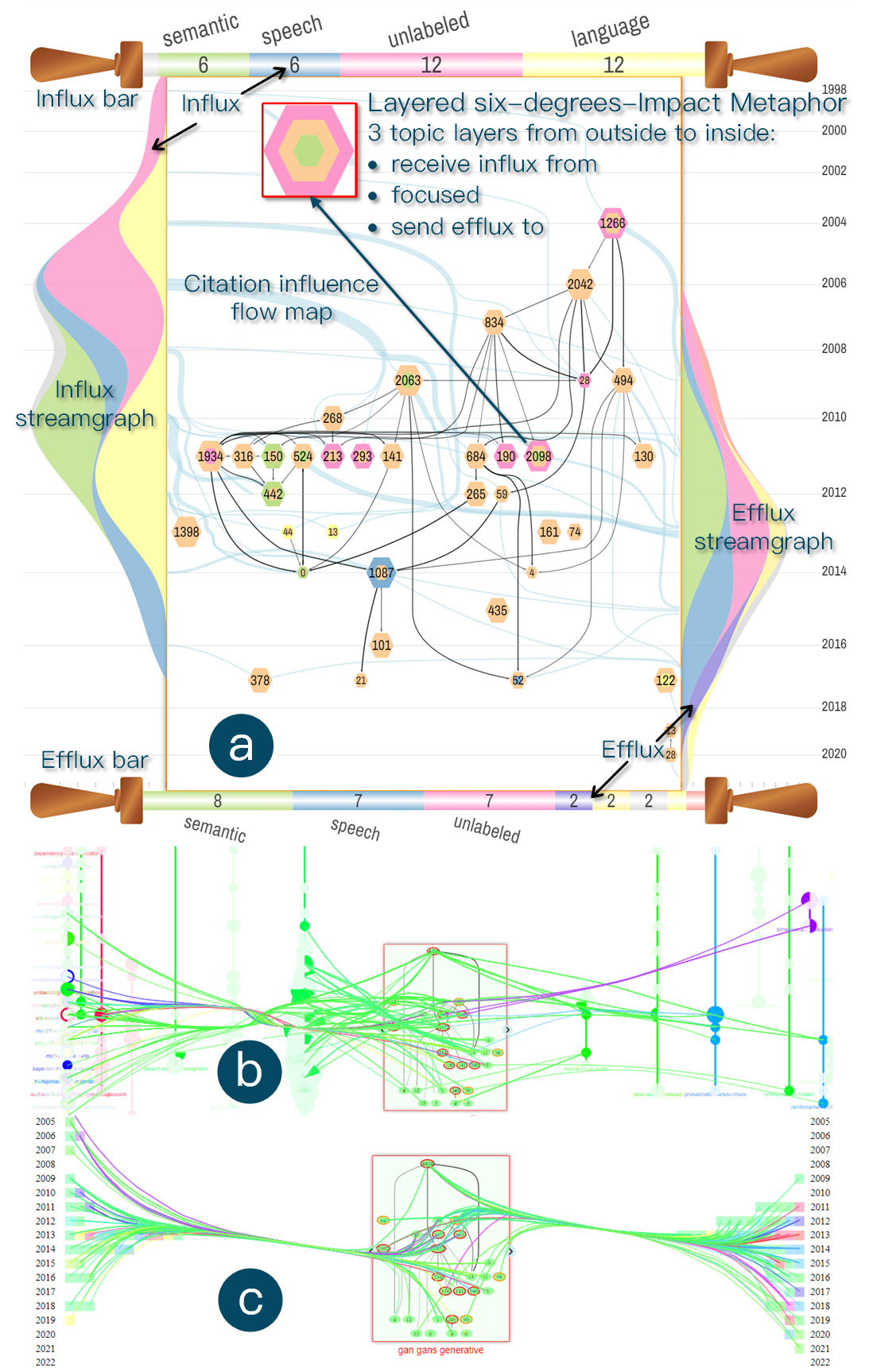}
  \vspace{-1.5em}
  \caption{%
  	GeneticScroll design and its alternatives. (a) GeneticScroll composed of a GF sub-graph visualization in the center, influx/efflux streamgraphs on the left/right sides, and citation influence flow maps in between showing detailed inter-topic interactions; (b)(c) two alternative designs.\shil{This figure's annotations should be updated according to the new text in the paragraphs below.}\sunye{fixed}
  }
  \label{fig:GeneticScroll}
  \vspace{-1em}
\end{figure}


The main panel of our system will switch to the GeneticScroll view when the GF sub-graph in black curved lines of one topic is focused on the default prism metaphor. As shown in \rfig{GeneticScroll}(a), the GeneticScroll design consists of a hierarchical visualization of GF sub-graph for the detailed analysis of a scholar's impact evolution on a single topic (T2), with elaborate glyph designs on graph nodes to represent interdisciplinary papers (T4); two streamgraphs in the side view to display the topic dynamics influencing and influenced by the current topic; and the flow map in blue curved lines connecting between the streamgraphs and the GF sub-graph to illustrate detailed topic interaction patterns surrounding the current topic (T3). The overall GeneticScroll design resembles a traditional scroll painting, highlighting the detailed and unfolding narrative of a scholar's scientific impact on a specific topic and its outreach.


The main component of GeneticScroll features the GF sub-graph visualization on the current topic, using the hierarchical layout with a fixed time axis similar to the side view of GeneticPrism. Since GeneticScroll is an expanded version of the side view, two design improvements are introduced to support an in-depth analysis of impact evolution on that topic. First, each paper node is now represented by a new glyph design of a layered six-degree-impact metaphor to highlight interdisciplinary papers. Second, the hierarchical node layout is optimized by a new algorithm to strike a balance between the layout athletics and the additional flow map visualization showing topic-level citation influences. The algorithm will be described in \rsec{Algorithm}.

\textbf{Layered six-degree-impact metaphor} for paper nodes, as enlarged in \rfig{GeneticScroll}(a), has a hexagon shape with the entire size visually encoding the class of high/medium/low citation number of the corresponding paper. The paper with more citations will be drawn in a larger hexagon. The default thresholds between high/medium/low citation classes are 100 and 50, which can also be tuned per the data set. In its full version, the hexagon glyph has three layers, each representing a topic to which the current paper belongs. By this design, the paper with overlapping topics, i.e., the interdisciplinary work, can be visually highlighted. 

In detail, the middle layer of the glyph represents the focused topic of GeneticScroll, e.g., yellow in \rfig{GeneticScroll}(a). The outermost layer corresponds to another topic to which the current paper not only belongs but also receives an influx of citation influence from papers on that topic, e.g., pink in \rfig{GeneticScroll}(a). On the visualization, there should be at least one flow map link from the left-side streamgraph layer of that topic to the current paper node. In case there are multiple such topics, the topic with the highest similarity to the current paper is used. Similarly, the innermost glyph layer corresponds to the topic to which the current paper is sending efflux citation influence, e.g., green in \rfig{GeneticScroll}(a). As not all papers are necessarily interdisciplinary, the glyph degenerates into a one-layer or two-layer hexagon in most cases. The size of multiple layers on a single paper hexagon is determined in proportion to the paper's similarity to these topics, as calculated in \rsubsec{Topic}. Note that we do not display the external topic on a paper having no interaction with the current topic/paper because this information will be visualized separately in the GeneticScroll view of that external topic.



\textbf{Influx/efflux streamgraphs} are presented on the left/right side of the GeneticScroll view, which displays the temporal dynamics of influx/efflux citation influence between other topics and the focused topic shown in GeneticScroll. Take the influx streamgraph in the left as an example, it is composed of multiple layers each corresponding to a topic sending citation influence to the current topic. The height of each layer at a year indicates the total amount of influx citation influence from that topic to the focused topic in GeneticScroll in that year. Topic colors inherit those of GeneticPrism and GeneticScroll. In addition, on the top/bottom of GeneticScroll, horizontal influx/efflux bars are designed to provide an overview of citation influence between external and focused topics. Each bar on the top/bottom indicates the influx/efflux of a topic, with the bar length proportional to the aggregated influx/efflux between that topic and the focused topic.


\textbf{Citation influence flow map} between influx/efflux streamgraphs and the GF sub-graph visualization in the center unveils the detailed topic-level interaction. The raw data are the citation influence links between the papers displayed in the current GF sub-graph and the papers on the other topics. Depicting these influence links altogether will lead to substantial visual clutter due to the GF sub-graph already presented. We propose three approaches to resolve this issue. First, the raw citation influence links are aggregated into influx/efflux flows by the year granularity and connected to the edges of left/right-side streamgraphs. Flow thickness will indicate the strength of that flow. Second, the classical flow map layout \cite{phan2005flow} is adopted to arrange the flow visualization appropriately. We introduce a new algorithm in \rsec{Algorithm} to optimize the layout with the best effort and minimize edge crossings. Third, the influx/efflux flow map is drawn by a light blue color in the background layer, distinguished from the GF sub-graph links drawn by the black color in the foreground layer.



GeneticScroll visualization also supports several interactions to speed up the visual analysis. In addition to pan\&zoom, clicking on nodes and links of the GF graph unfolds the details of selected papers and citation influence links in the information panel of \rfig{GeneticPrismSystem}(e), including the paper title, abstract, venue, year, topics, and citation context, etc. Hovering an influx/efflux streamgraph layer highlights the part of the flow map representing the influx/efflux flows from/to the corresponding topic. Many other visualization parameters can also be visually configured, such as the year granularity for aggregating flows, the weight to layout GF graph vs. flow map, and the mode of edge bundling.


\textbf{Alternative design} of GeneticScroll includes:
\begin{itemize}
    \item \emph{GF sub-graph visualization by gourds (\rfig{GeneticScroll}(b))} displays the GF sub-graph of a topic at the center of the interface. Papers in the other topics are arranged as topic strings in a gourd-like shape surrounding the central GF sub-graph on the left and right. Each gourd in a string is drawn by a pie chart for the paper set in a particular year, with citation influence links connected to the central nodes. The resulting visualization is very cluttered, with severe edge crossings.
    \item \emph{GF sub-graph visualization by stacked bar charts (\rfig{GeneticScroll}(c))} replaces the streamgraph design in GeneticScroll with stacked bar charts, but the temporal dynamics of these related topics are hardly perceived.
\end{itemize}

\section{Integrated Flow Hierarchical Layout}
\label{sec:Algorithm}

\begin{figure}[tb]
  \centering
  \includegraphics[width=\columnwidth, clip, trim=10pt 10pt 10pt 5pt]{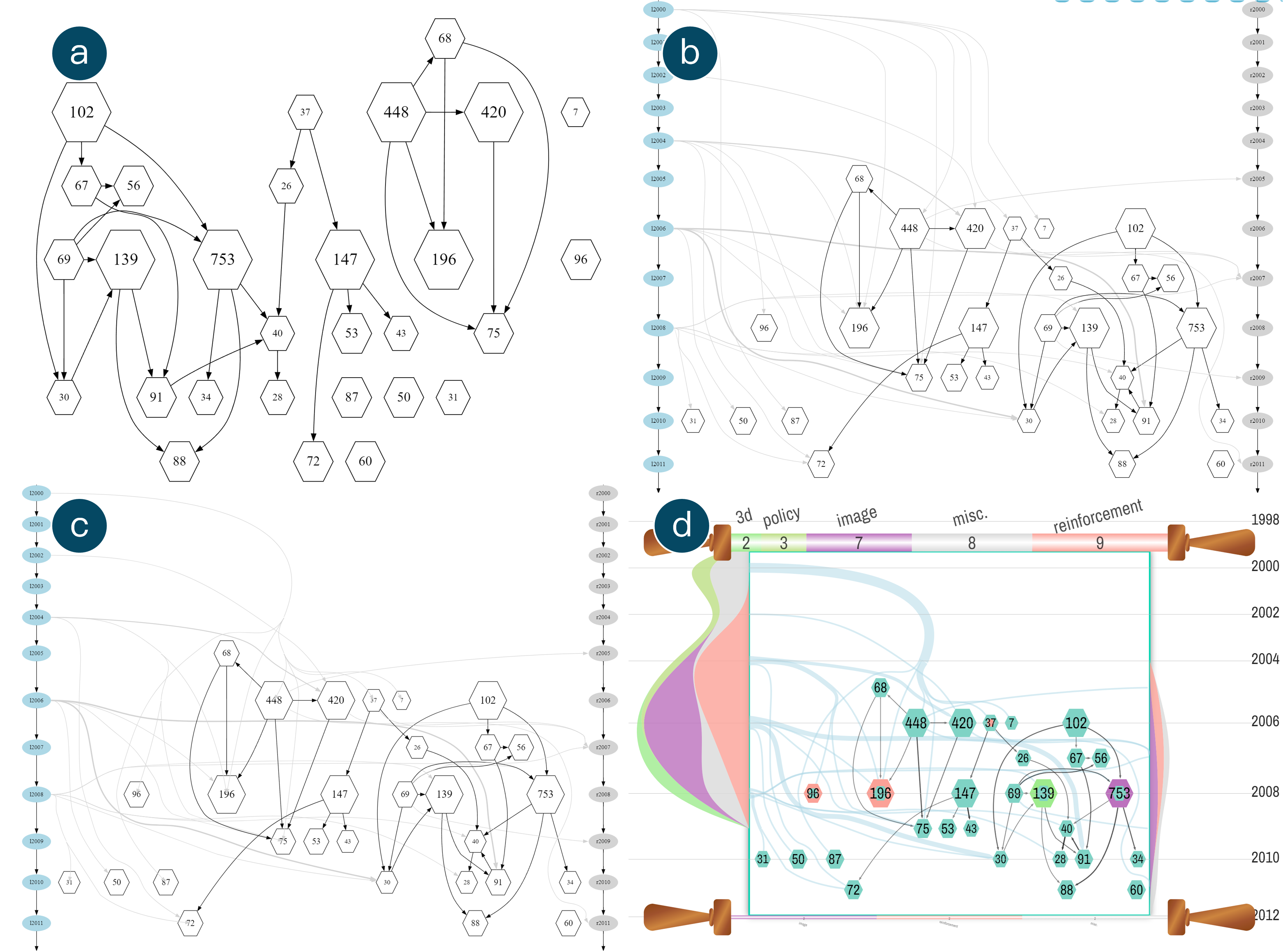}
  \vspace{-1.5em}
  \caption{%
  	Steps from the original GF sub-graph layout to the GeneticScroll view. (a) Original per-topic GF sub-graph; (b) GF sub-graph with influence edges; (c) GF sub-graph with influence edges using bundling; (d) GeneticScroll view: add flow map layout for influence edges and place them on the background layer, integrating with influx/efflux streamgraphs.%
  }
  \label{fig:HierarchicalLayout}
  \vspace{-1em}
\end{figure}

We propose the Integrated Flow Hierarchical Layout (IFHL) algorithm for the GeneticScroll visualization, considering not only the layout of the central graph but also integrating its influence on other topics through flow map and streamgraph. In contrast to the designs in \rfig{GeneticScroll}(b) and \rfig{GeneticScroll}(c), which directly draw citation influences over the per-topic GF sub-graph, our approach integrates the layout of the central graph and citation influences together. As shown in \rfig{HierarchicalLayout}, three steps are adopted to distinguish the central graph from the newly added influence flow map and effectively prevent visual clutter.

The objective of IFHL is to lay out the context edges integrated with the original hierarchical layout. The set of context edges includes all edges that are connected directly to the topic \( T_k \) and that are not within \( T_k \), defined as:

\vspace{-0.5em}
\begin{equation}
\begin{aligned}
    E_C(T_k) = \{ e_{ij} \mid e_{ij} \in E, &v_i \notin V(T_k) \wedge v_j \in V(T_k) \\
    \vee &v_i \in V(T_k) \wedge v_j \notin V(T_k) \}
\end{aligned}
\label{eq:context}
\end{equation}

The ``in'' edges, or influx, connect to the upstream of the topic, while the ``out'' edges, or efflux, connect to those downstream. To display the influx and efflux patterns, we place the ``in'' and ``out'' edges on the left and right sides, respectively. We introduce influx nodes \( v_l(y) \) and efflux nodes \( v_r(y) \), and \( y \) represents the year associated with the influx/efflux node. These nodes act as proxies for the endpoints of context edges. If a context edge connects from a node outside the topic sub-graph to a node inside, the source of the edge is replaced by the corresponding influx node. Similar to the efflux node. By replacing the original endpoint of the context edges with influx/efflux nodes, we get influence edges, defined as:

\vspace{-1em}
\begin{equation}
\begin{aligned}
    E_I(T_k) &= \{ e_{pj} \mid e_{ij} \in E_C(T_k), v_j \in V(T_k) \rightarrow v_p = v_l(\text{year}(v_i)) \} \\
\cup &\{ e_{ip} \mid e_{ij} \in E_C(T_k), v_i \in V(T_k) \rightarrow v_p = v_r(\text{year}(v_j)) \}
\end{aligned}
\label{eq:proxy_graph}
\end{equation}

During the layout of the GF graph, we adhere to the rules that influx/efflux nodes are always positioned on the left and right sides, arranged by year. We enhanced the Hierarchical Layout provided by GraphViz (DOT) to lay out the whole graph with influence edges and influx/efflux nodes rather than the central graph. During the node ordering step, we ensure that influx nodes \( v_l \) and efflux nodes \( v_r \) are positioned at the extreme left and right of the layout, respectively.

\textbf{Weighted edge crossing} is adopted as the optimization objective rather than edge crossing. The weight of edges in the central GF graph is set to \( \alpha \), and the weight of influence edges on the flow map is proportional to edge width so that strong influences show more importance in the layout. In the objective function, we use \( \text{cost}(e) \) instead of the default ``1'' as the edge weight, defined as:

\begin{equation}
\begin{aligned}
    \text{cost}(e) &=
    \begin{cases}
        \alpha & \text{if } e \in E(T_k) \\
        \text{weight}(e) & \text{if } e \in E_I(T_k)
    \end{cases}
    \label{eq:crossing}
\end{aligned}
\end{equation}

The weighted crossing between \( e_1 \) and \( e_2 \) is defined as \( \text{cost}(e_1) \cdot \text{cost}(e_2) \). We use total weighted crossing as the optimization objective. To evaluate the optimal weight \( \alpha \) for the best layout, we use crossing counts as indicators, including internal crossing (between central edges) and external crossing (between influence edges). We aim to find an optimal objective where the number of internal crossings is low (primary goal) while maintaining low external crossings. Finally, the elbow method is introduced to determine the optimal \( \alpha \). 

In summary, we select \( \alpha \) incrementally, stopping when the reduction in internal crossing does not outweigh the increase in external crossing. Our algorithm calculates an adaptive \( \alpha \), resulting in an enhanced topic GF graph with detailed influence information and minimal deviation from the original layout. Detailed steps and the algorithm for these processes are provided in the appendix.

\textbf{Citation influence flow map} in GeneticScroll ensures that edge widths increase progressively at intersection points, similar to a Sankey diagram, to prevent visual clutter and edge crossings (\rfig{HierarchicalLayout}(d)). We achieve this through a two-step process: edge bundling and flow map adjustment. The method results in a clear and comprehensible visualization of complex network flows, effectively integrating the central graph's layout and influence edges.

In the edge bundling step, we merge adjacent dummy nodes into a single node, focusing on influence edges while keeping the central graph unchanged (\rfig{HierarchicalLayout}(c)). This helps reduce edge crossings and make the graph more compact. By treating central edges and influence edges separately, we ensure that the central graph's pattern remains clear. Only influence edges (\( v_{cc} \) or \( v_{cp} \)) are bundled to simplify visual elements and highlight the topic GF graph.

The flow map adjustment step refines the layout based on the bundled edges (\rfig{HierarchicalLayout}(d)). We model intersection points and adjust positions, thicknesses, and order of split edges to achieve a Sankey diagram style. This involves constructing an intersection tree from the pre-arranged Bézier curves, performing a topological sort to ensure progressive edge widths, and adjusting upstream curves. Detailed steps and the algorithm for flow map adjustment are provided in the appendix.

\section{Case Studies and Evaluation}
\label{sec:Case}
\subsection{Prof. Hanrahan's Interdisciplinary Impact}

\begin{figure*}[tb]
  \centering
  \includegraphics[width=\textwidth, clip, trim=3pt 0pt 0pt 16pt]{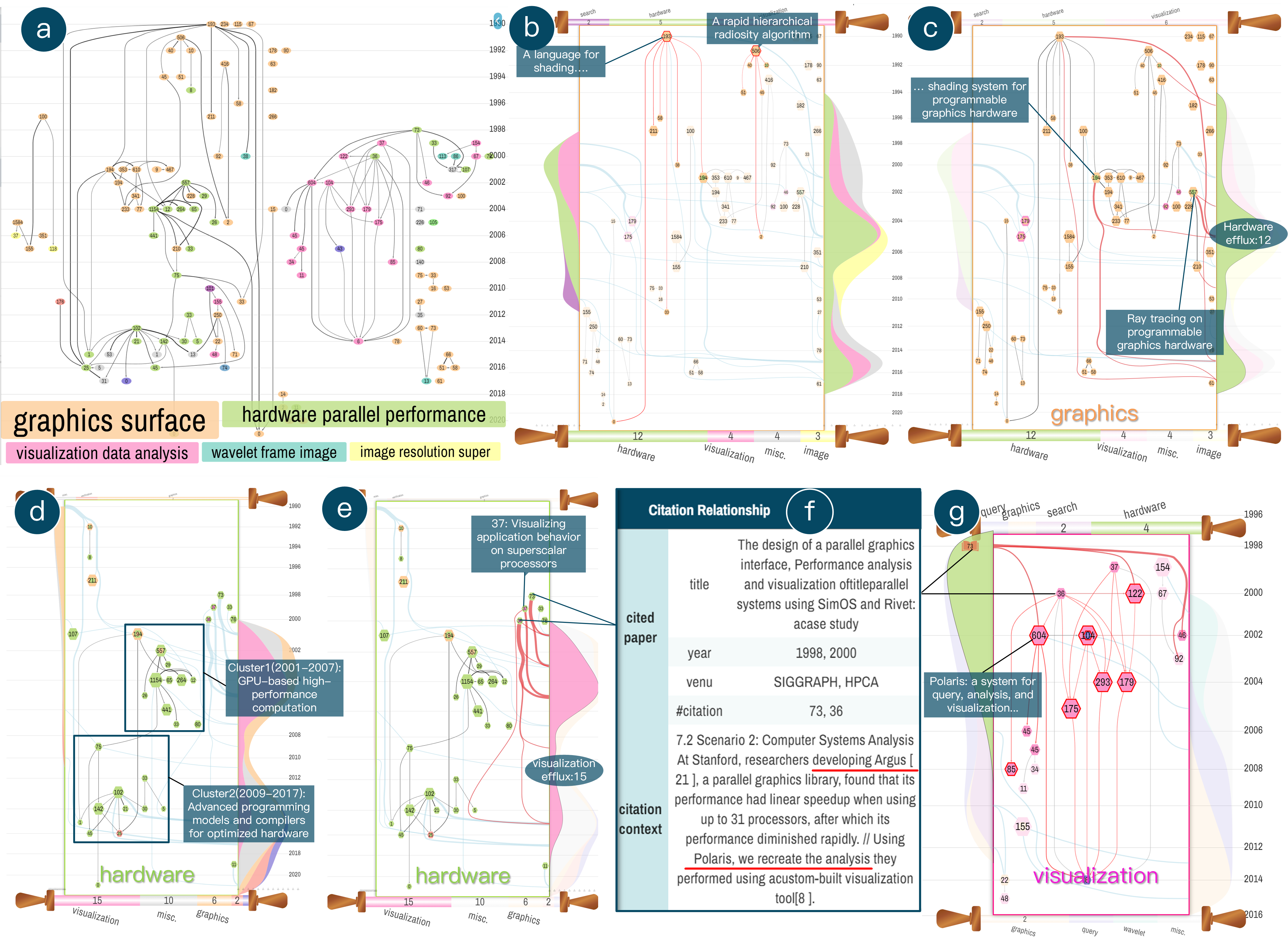}
  \vspace{-1.8em}
  \caption{%
  	Case study on Pat Hanrahan's interdisciplinary impact: (a) full GF graph; (b,c) GeneticScroll view on the topic ``3D graphics''; (d,e) GeneticScroll view of topic ``parallel hardware''; (f) detailed citation information of paper ``Polaris''; (g) GeneticScroll view on the topic ``visualization''.%
  }
  \label{fig:case1}
  \vspace{-1em}
\end{figure*}


Prof. Pat Hanrahan is an ACM Turing award laureate (2019) known for fundamental contributions to 3D computer graphics and computer-generated imagery \cite{HanrahanTuring}. 
The first case study leverages GeneticPrism to visually analyze Hanrahan's scientific impact evolution, focusing on his contributions to interdisciplinary research topics.

Initially, using the original GF graph visualization (\rfig{case1}(a)), we identify three primary research topics associated with Hanrahan's work: 3D graphics (orange), parallel hardware (green), and visualization (pink). However, the three topics somehow intertwine over time, and it is challenging to see the evolution structure of each topic clearly. Though the topic interaction can be observed through links between differently colored nodes, since each paper is only encoded as one topic, it is also impossible to identify key interdisciplinary works and their impact enclosure.

By the new GeneticScroll view, Prof. Hanrahan's scientific impact on the 3D graphics topic over 30 years can be summarized as a single picture of \rfig{case1}(b). Notably, we can identify the two pioneering papers with red outlines on this topic. The first is the RenderMan shading language paper published in 1990 (193 citations), and the second developed the hierarchical radiosity algorithm in 1991 (506 citations). The following of Hanrahan's 3D graphics works mainly extend from these papers into two research threads: one focuses on shading language and the other on textures and radiosity.

From the early 1990s, as his Turing award statement also mentions, the 3D graphics topic gradually incubated another topic alongside, as evidenced by the green layer in the right-side streamgraph (\rfig{case1}(c)). This is the parallel hardware topic, which mainly refers to the use of GPUs to accelerate graphics processing. The total efflux from 3D graphics to parallel hardware amounts to 12 extensions. The topic interactions are featured by two influential, interdisciplinary papers tagged in \rfig{case1}(c), all having colored layers on both topics. The first SIGGRAPH'01 paper (194 citations) describes a shading system on new graphics hardware, and the second SIGGRAPH'02 paper (577 citations) deploys ray tracing algorithms to the GPU hardware.

\rfig{case1}(d) gives an overview of Hanrahan's work on the parallel hardware topic. His contributions there can be roughly clustered into two sub-topics: GPU-based high-performance computation for graphics and advanced graphic programming models and compilers.\sunye{Removed: graphics language/ray tracing on new hardware, and the DSL compiling research} \shil{What is DSL?} Interestingly, the topic again triggers a new topic on visualization, as shown by the pink-layer streamgraph in \rfig{case1}(e), with a topic efflux of 15 extensions. Three papers close to the pink layer contribute the most to this transition of research. Two of these papers are interdisciplinary between parallel hardware and visualization, as shown by the multiple color layers. After mouse hovers to access their titles, these researches are found to be the visualization of parallel hardware's performance and application behavior. We further drill down to the citation influence links from the hardware research to visualization. The citation contexts in \rfig{case1}(f) reveal a key clue on how they developed the famous Polaris visualization system: researchers first invented Argus \cite{igehy1998design} (a parallel graphics library, one of the three papers, 73 citations), but found scalability issues on its performance, so they decided to apply visualization tools to analyze the performance \cite{bosch2000performance} (another of the three papers, 36 citations).

In \rfig{case1}(g), the evolution of Hanrahan's visualization work is illustrated. By examining all paper titles, it can be found that the primary component is seven publications focusing on the Polaris visualization system, with three papers, three patents, and one case study about the system, as indicated by the red outline. Notably, one KDD'02 paper \cite{stolte2002query} within them is shown to be interdisciplinary with database research. A further drill-down into the details explains this finding as the application of Polaris to the analysis of multidimensional databases. It is indeed well-known that Polaris/Tableau's initial customers are from the database community.

In a few visual analysis walks with GeneticPrism, we unveil the three major contribution areas of Prof. Hanrahan, as well as his proven stories on topic evolution and transition. The case strongly demonstrates the effectiveness of our visualization technique in illustrating and explaining both intra-topic and inter-topic scientific impact evolution on individual scholars.

\subsection{Prof. Stonebraker's Lifelong Database Research}

\begin{figure}[tb]
  \centering
  \includegraphics[width=\columnwidth, clip, trim=10pt 30pt 10pt 10pt]{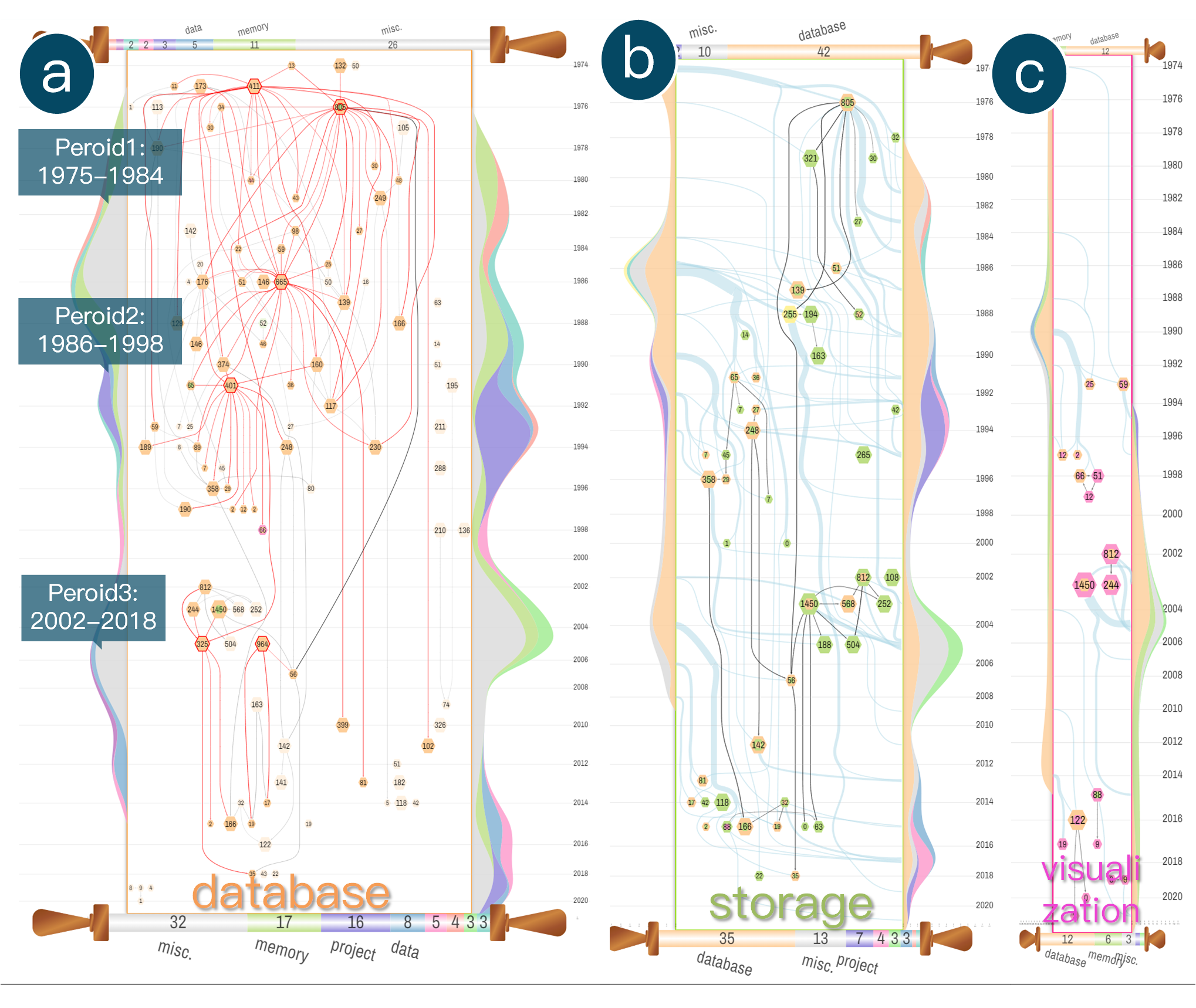}
  \vspace{-1.5em}
  \caption{%
  	Case study on Michael Stonebraker's academic career: (a)(b)(c) GeneticScroll visualization of ``database'', ``storage'', and ``visualization'' topics, respectively.%
  }
  \label{fig:case2}
  \vspace{-1em}
\end{figure}

In the second case study, we apply GeneticPrism to understand the research impact of Prof. Michael Stonebraker, another renowned ACM Turing award laureate (2014). Prof. Stonebraker has significantly influenced the field of databases, and his pioneering works have laid the foundation for many modern database technologies.

\rfig{GeneticPrismSystem} provides a topic overview of Prof. Stonebraker's research by the chord diagram in GeneticPrism. We observe two primary research topics: database (orange) and storage (green), along with several minor research topics such as visualization (pink) and cloud. The database topic dominates his research with 107 core papers out of all 174 papers (61.5\%). The GeneticPrism view (\rfig{GeneticPrismSystem}(d)) putting his two major topics side by side echoes the dominance of the database topic over the entire period and exposes the richness and connectedness of his research on the topic. In comparison, the storage topic is featured only in a few research threads with tens of publications.

Drilling down to the GeneticScroll panel for details on the database topic, \rfig{case2}(a) sketches a global view showing densely interrelated research. The hierarchical node-link graph drawing attached to the time axis identifies three major periods of Prof. Stonebraker's database research, each with a few landmark papers influencing tens of follow-up papers to form a thread. In the first period (1975-1984), two papers with 411 and 805 citations (in red outline) introduce the INGRES system. In the second period (1986-1998), the two papers with 665 and 401 citations developed the Postgres system. The third period (2002-2018) features two papers with 325 and 964 citations, both studying the latest commercial Database Management System (DBMS). These discoveries correspond precisely to the three contributions in Prof. Stonebraker's Turing award statement \cite{StonebrakerTuring}: INGRES, Postgres, and Entrepreneurship on DBMS. By fixing these six key papers in the figure, their influences almost cover the entire landscape of his database research.

We further expand the GeneticScroll views on his ``storage'' and ``visualization'' topics (\rfig{case2}(b) and \rfig{case2}(c)). Unlike the database view, the common characteristic of the two topics lies in that both figures are not overwhelmed by the works from the current topic facet. Instead, paper nodes (partly) in orange appear everywhere on the two topic graphs, indicating interdisciplinary research with the database. The same pattern can be observed in other minor topic facets of Prof. Stonebraker. We infer that most of his works on other topics are about applications of database technology or tool support for database systems (e.g., visualization). GeneticPrism visualizations effectively delineate Prof. Stonebraker's lifelong dedication to database research.

\subsection{The Backbone and Outreach of Graph Drawing Symposium}

\begin{figure}[tb]
  \centering
  \includegraphics[width=\columnwidth, clip, trim=10pt 30pt 10pt 5pt]{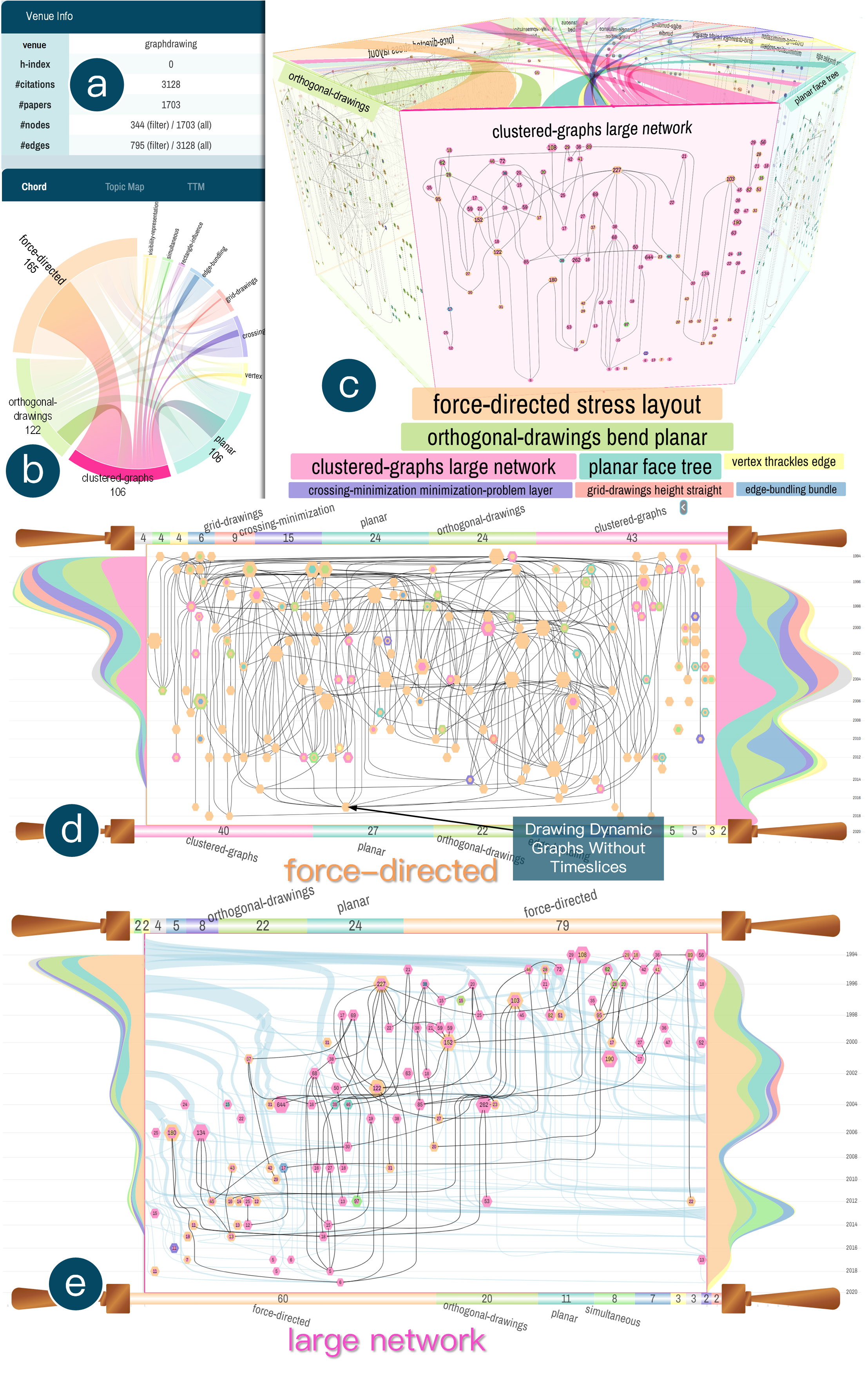}
  \vspace{-1.5em}
  \caption{%
  	Case study on the GD symposium: (a) scholar demographics; (b) chord diagram; (c) GeneticPrism visualization; (d)(e) GeneticScroll visualization of ``force-directed'', ``clustered-graphs'' topics, respectively.%
  }
  \label{fig:case3}
  \vspace{-1em}
\end{figure}

Beyond the above cases on prestigious scholars, the GeneticPrism technique can also be applied to understand the evolutionary impact of academic events, e.g., International Symposium on Graph Drawing and Network Visualization (GD)~\cite{GD}. We extract all the GD papers from MAG (1994$\sim$2021), up to 1703 papers. To accommodate this large data in GeneticPrism, only high-impact papers are presented, filtering by their number of citations: $\geq$ 5 for papers newer than 2017, $\geq$ 10 for papers newer than 2012, and $\geq$ 15 for other papers. The final GD dataset has 344 papers (nodes) and 795 citations (edges).

As illustrated in \rfig{case3}(b), these high-impact GD papers form 11 topics, with a single backbone topic of ``force-directed'' having 165 papers, covering half of all GD papers. Three other topics displayed in \rfig{case3}(c) are the second largest (also indicated by the arc thickness in \rfig{case3}(b)): ``orthogonal-drawings'' published before 2014 (with over 90\% papers), ``planar drawing and trees''\shil{should give the year bound}, and the ``clustered-graph and large network visualization'' topic studied in the whole time of GD.


Drilling down to the force-directed topic with GeneticScroll reveals its backbone nature for GD. In the densely connected graph of \rfig{case3}(d), more than half of all nodes are encoded with multiple color layers, representing interdisciplinary research applying force-directed methods to various graph drawing problems. Meanwhile, it is observed that many of the highest-impact papers drawn in the largest hexagons are single-colored. This suggests that the inherent advancement of force-directed methods still serves as a major scientific evolution in the GD symposium. 
The GeneticScroll view of the large network visualization topic with lasting impact presents a major outreach of GD (\rfig{case3}(e)). The topic was initially influenced by force-directed methods. After that, new requirements in visualizing large graphs in return drove the development of force-directed drawing, as shown by the streamgraph in the right part of \rfig{case3}(e). In the recent years of 2010$\sim$2016, it even triggered a new research topic on ``simultaneous embedding'' (in green color), which lays out multiple graphs simultaneously to solve the challenge of multi-level or multi-component drawing in visualizing large graphs.

\section{Discussion}


Through the case study, we demonstrate that GeneticPrism visualization can effectively delineate the major research threads of top scholars and academic venues and help analyze both intra-topic and inter-topic impact evolutions. Compared with previous visualization methods, the advantage lies in the strategy of slicing one's citation influence graph into different topic facets. This not only reduces the visual complexity in each single view but also allows us to visually analyze the impact evolution of each topic independently.


Meanwhile, a potential limitation of GeneticPrism is its customized design for the self-citation influence graph of an individual scholar. This positions the current work towards an application-oriented approach. Yet, in an effort to elicit generic methodology from the approach, it can be found that the same GeneticPrism design can be readily extended to visualize generic multivariate networks with a primary node attribute working as the topic attribute in GeneticPrism. The assumption would be that this node attribute is nominal/categorical, though careful pre-processing can also be introduced to accommodate numeric/ordinal node attributes. In this way, the GeneticPrism approach will work well on multivariate networks where nodes with the same attribute value stay more closely together than nodes with different values. These assumptions could be met frequently; in the extreme case, graph clustering can set up an applicable scenario when the cluster-ID works as the primary node attribute.


Another potential future work would be extending GeneticPrism to support the analysis of multiple scholars in the same group or community. The single hierarchy design from GeneticPrism to GeneticScroll can be upgraded to having multiple hierarchies, with the scholar facets at the top and more topic facets at the bottom. To reduce the visual complexity in the GF graph of each scholar, network abstraction/aggregation techniques should be studied carefully. Since the number of topics involving multiple scholars will increase a lot, hierarchical topic modeling can also be introduced to create multiple levels in the topic facet visualization of GeneticPrism.

\section{Conclusion}
This work studies an essential problem of visualizing the scientific impact evolution of individual scholars from multiple topic facets. Building on the previous work of GeneticFlow, we propose an end-to-end technical framework that includes academic data curation and pre-processing, citation influence graph analysis and topic modeling, and the GeneticPrism visualization designs and interactions, which are our main contributions. A new 3D prism-shaped visual metaphor is introduced for impact evolution overview, while a GeneticScroll design is proposed to orchestrate both intra-topic temporal/hierarchical impact evolution and inter-topic citation interactions, with an elaborate layout algorithm to minimize edge crossing and visual clutter. The visualization techniques are demonstrated to be effective through two case studies in delineating the intriguing academic lives of Turing award laureates, as well as an example scenario applying GeneticPrism to the GD symposium and analyzing the scientific evolution of its more than 1000 research papers.



%



\bibliographystyle{abbrv-doi-hyperref}

\bibliography{GeneticPrism}





\vspace{-2em}
\begin{IEEEbiography}[{\includegraphics[width=1in,height=1.25in,clip,keepaspectratio]{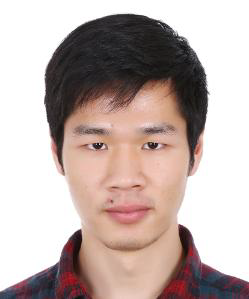}}]{Ye Sun}
is currently a Ph.D. student at the School of Computer Science at Beihang University. His main research interests include visualization and data mining.
\end{IEEEbiography}
\vspace{-2em}

\begin{IEEEbiography}[{\includegraphics[width=1in,height=1.25in,clip,keepaspectratio]{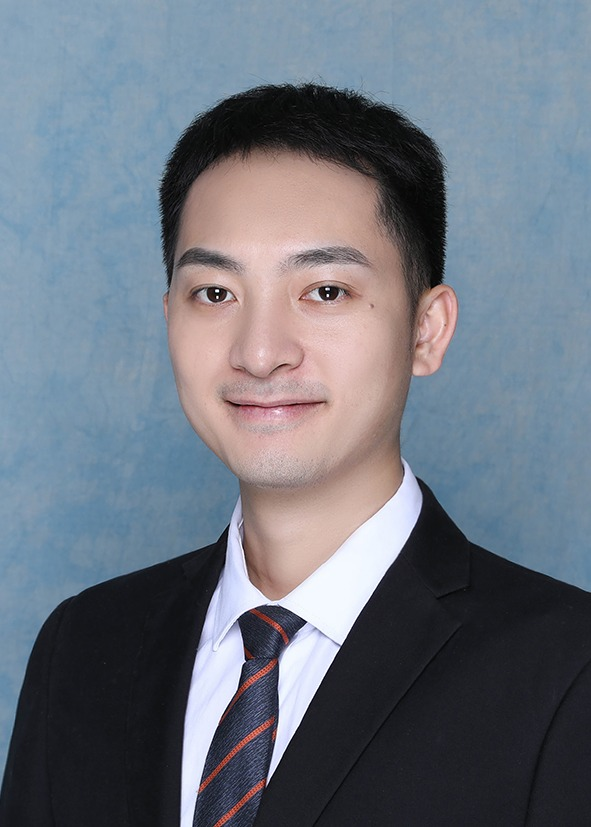}}]{Zipeng Liu}
is an Assistant Professor at the School of Software, Beihang University. He received his Ph.D. in Computer Science from the University of British Columbia, Canada, in 2021. His primary research interests include visualization and visual analytics, human-computer interaction, big data analysis, and explainable machine learning. He has been recognized with a Best Paper Nomination Award at PacificVis. He also serves as a reviewer for several top-tier academic journals and conferences.
\end{IEEEbiography}
\vspace{-2em}

\begin{IEEEbiography}[{\includegraphics[width=1in,height=1.25in,clip,keepaspectratio]{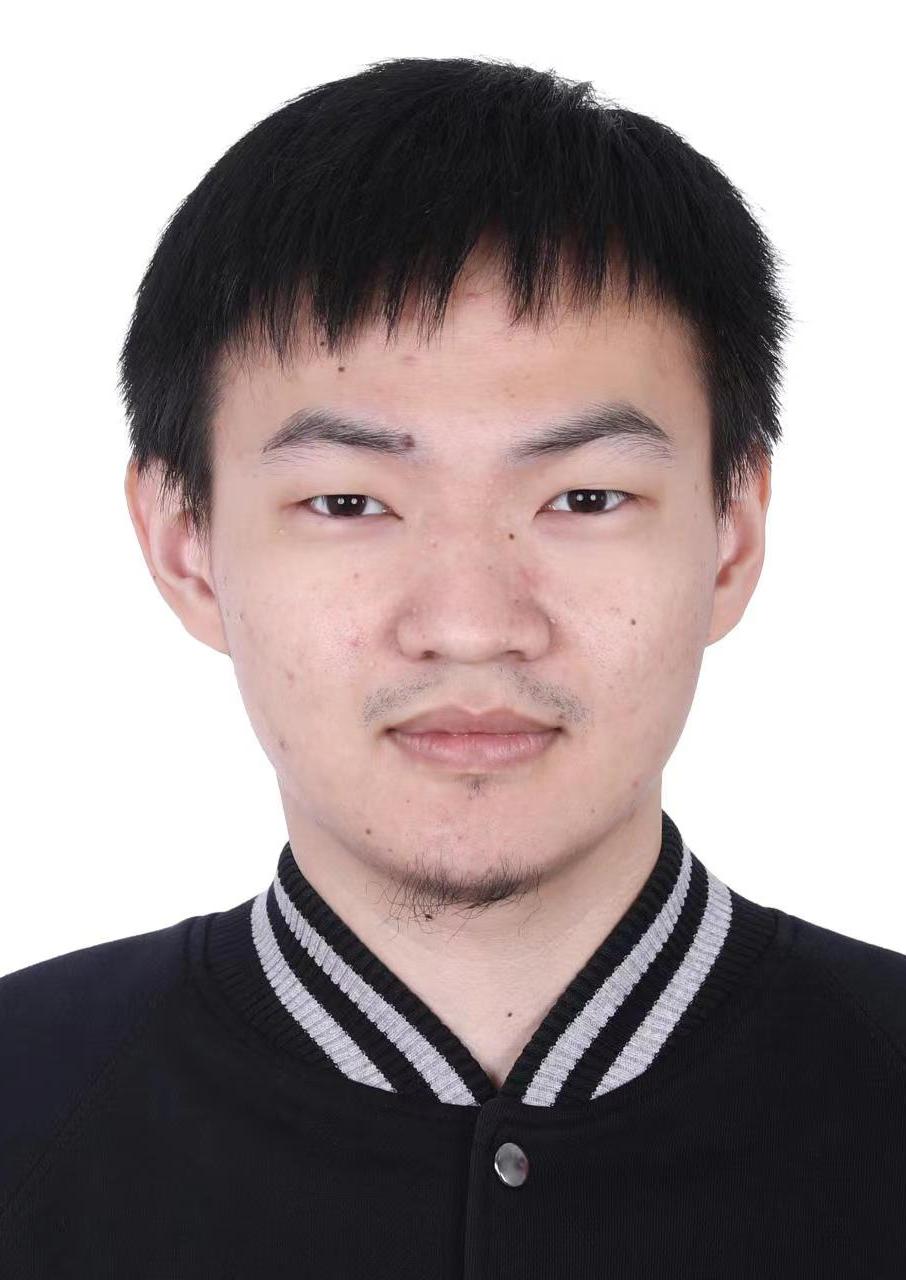}}]{Yuankai Luo}
is a Ph.D. candidate in the School of Computer Science at Beihang University. His primary research interests are in data mining and graph learning. He has served as a reviewer for several top-tier academic journals and conferences, including TKDE, TNNLS, ICML, and NeurIPS.
\end{IEEEbiography}
\vspace{-2em}

\begin{IEEEbiography}[{\includegraphics[width=1in,height=1.25in,clip,keepaspectratio]{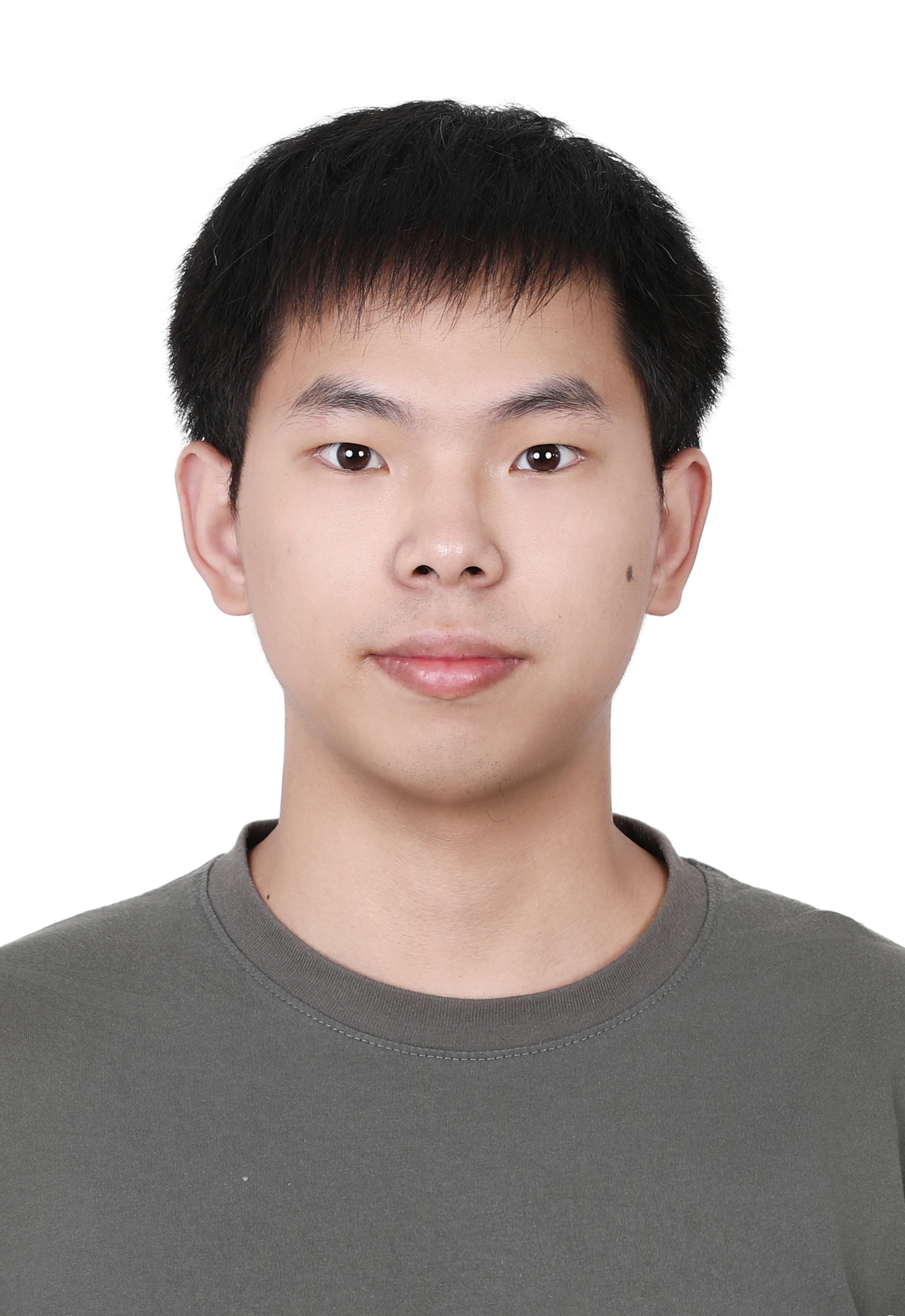}}]{Lei Xia}
is currently a Ph.D. student at the School of Computer Science at Beihang University. His main research interests include data mining and graph analysis.
\end{IEEEbiography}
\vspace{-2em}

\begin{IEEEbiography}[{\includegraphics[width=1in,height=1.25in,clip,keepaspectratio]{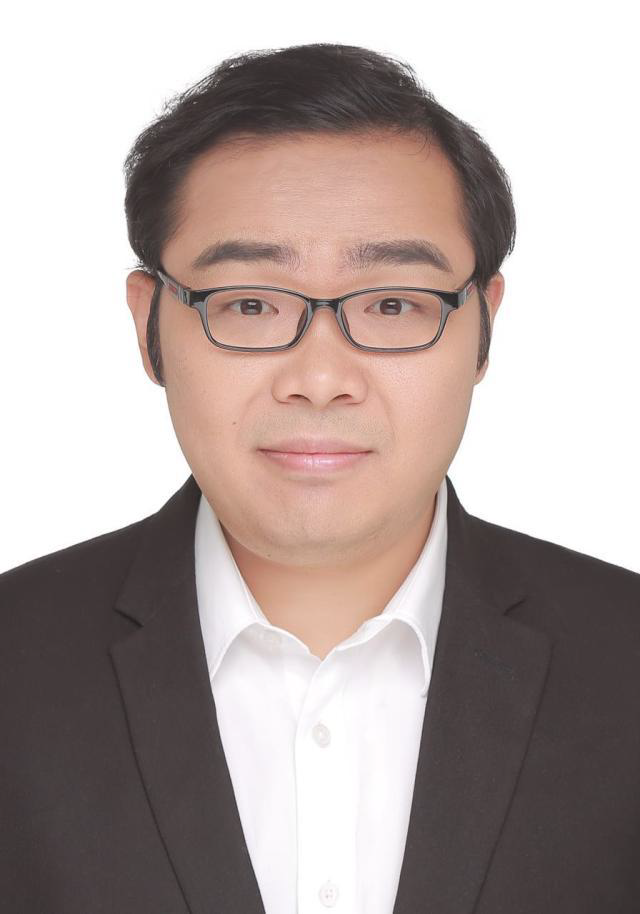}}]{Lei Shi}
is currently a professor at the School of Computer Science and Engineering at Beihang University. He holds B.S., M.S., and Ph.D. degrees from the Computer Science and Technology Department at Tsinghua University. His current research interests are visual analytics, data mining, and AI, with more than 100 papers published in international conferences and journals. He received the IBM Research Division award on visual analytics and the IEEE VAST Challenge award twice in 2010 and 2012. He is an IEEE senior member.
\end{IEEEbiography}


\vfill

\end{document}


\maketitle

\section{Integrated Flow Hierarchical Layout}
\begin{figure}[tb]
  \centering 
  \includegraphics[width=\columnwidth, clip, trim=10pt 10pt 10pt 10pt]{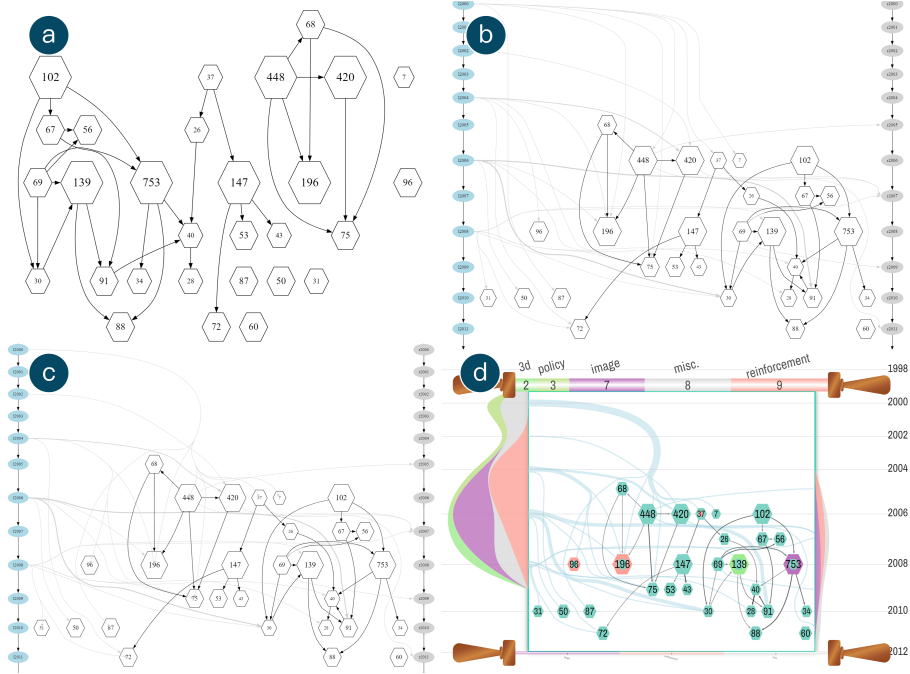}
  \caption{%
  	Steps from the original GF sub-graph layout to the GeneticScroll view. (a) Original per-topic GF sub-graph; (b) GF sub-graph with influence edges; (c) GF sub-graph with influence edges using bundling; (d) GeneticScroll view: add flow map layout for influence edges and place them on the background layer, integrating with influx/efflux streamgraphs%
  }
  \label{fig:HierarchicalLayout}
\end{figure}

In GeneticPrism visualization, we utilize the hierarchical layout algorithm to visualize the prism's per-topic GF sub-graph on each lateral face. However, it overlooks the interactions between the central and related topics. To address this, we propose the Integrated Flow Hierarchical Layout (IFHL) for the GeneticScroll visualization, considering not only the layout of the central graph but also integrating its influence on other topics through the flow map and streamgraph. In contrast to alternative designs of GeneticScroll, which directly draw influence edges after drawing the topic GF sub-graph, our approach integrates the layout of the central graph and influence edges together. As shown in \rfig{HierarchicalLayout}(a) to \rfig{HierarchicalLayout}(d), three steps are adopted to distinguish the central graph from the newly added influences and effectively prevent visual clutter.

\subsection{Objective and Constraint}

To support the need for overlapping topics, we conduct GF graph partitioning for each topic. We construct a corresponding GF sub-graph for each topic, defined as:

\begin{equation}
\begin{aligned}
    G(T_k) &= \{V(T_k), E(T_k)\} \\
    V(T_k) &= \{ n \mid n \in V, T_k \in \text{topic}(n) \} \\
    E(T_k) &= \{ e_{ij} \mid e_{ij} \in E, v_i \in V(T_k) \wedge v_j \in V(T_k) \} 
\end{aligned}
\label{eq:topic_graph}
\end{equation}

In this formula, \(e_{ij}\) is an extension edge (a reverse citation edge), \(E\) is the set of all extension edges in the author(\(s\))'s collection, \(v_i\) and \(v_j\) are the source and target of edge \(e_{ij}\). The function \(\text{topic}\) maps a paper to its corresponding set of topics (with a maximum size of 3).

The set of context edges, denoted by Equation \eqref{eq:context}, includes all edges that are connected directly to the topic \( T_k \) and that are not within \( T_k \). The ``in'' edges, or influx, connect to the neighbor nodes upstream of the topic, while the ``out'' edges, or efflux, connect to those downstream.

\begin{equation}
\begin{aligned}
    E_C(T_k) = \{ e_{ij} \mid e_{ij} \in E, &v_i \notin V(T_k) \wedge v_j \in V(T_k) \\
    \vee &v_i \in V(T_k) \wedge v_j \notin V(T_k) \}
\end{aligned}
\label{eq:context}
\end{equation}

To display the influx and efflux patterns, we place the ``in'' and ``out'' edges on the left and right sides, respectively. We introduce influx nodes \( v_l(y) \) and efflux nodes \( v_r(y) \), and \( y \) represents the year associated with the influx/efflux node. These nodes act as proxies for the endpoints of context edges. If a context edge connects from a node outside the topic sub-graph to a node inside, the source of the edge is replaced by the corresponding influx node. Similar to the efflux node. By replacing the original endpoint of the context edges with influx/efflux nodes, we get the set of influence edges, defined as:

\begin{equation}
\begin{aligned}
    E_I(T_k) &= \{ e_{pj} \mid e_{ij} \in E_C(T_k), v_j \in V(T_k) \rightarrow v_p = v_l(\text{year}_l(v_i)) \} \\
\cup &\{ e_{ip} \mid e_{ij} \in E_C(T_k), v_i \in V(T_k) \rightarrow v_p = v_r(\text{year}_r(v_j)) \}
\end{aligned}
\label{eq:proxy_graph}
\end{equation}

Here, \( V(T_k) \) is the set of central nodes, \( E(T_k) \) is the set of central edges, \( E_C(T_k) \) is the set of context edges, and \( E_I(T_k) \) is the set of influence edges. The functions source and target map edges to their start and end points, respectively, and $\text{year}_l$ and $\text{year}_r$ map nodes to the years of the influx/efflux nodes. To further abstract and reduce visual elements, we aggregate years by granularity. The influx nodes are mapped to the earliest year within the granularity, and the efflux nodes are mapped to the latest year (given by Equation \eqref{eq:year}), ensuring a top-down extension. For example, for the year 2008, we create nodes \( v_l(2008) \) and \( v_r(2008) \). \( v_l(2008) \) represents all nodes entering the central graph from 2008, and all influx influence edges \( e_{ij} \) with \(\text{year}_l(v_k)=2008\) originate from \( v_l(2008) \). We compress context information using influx/efflux nodes into a single point, reducing visual elements.

\begin{equation}
\begin{aligned}
    \text{year}_l(v_k) &= \left\lfloor \frac{\text{year}(v_k)}{\text{grid}} \right\rfloor \cdot \text{grid} \\
    \text{year}_r(v_k) &= \left(\left\lfloor \frac{\text{year}(v_k)}{\text{grid}} \right\rfloor+1\right) \cdot \text{grid} - 1
\end{aligned}
\label{eq:year}
\end{equation}


\begin{table}[tb]
  \caption{%
    Description of Nodes, Edges, and crossings.%
  }
  \label{tab:nodes_edges_crossings}
  \scriptsize%
  \centering%
  \renewcommand{\arraystretch}{1.3}
  \begin{tabu} to \textwidth {%
        X[0.4,c]%
        X[0.3,c]%
        X[0.4,c]%
        X[2.2,l]%
        X[1.2,c]%
      }
    \toprule
    \textbf{Type} & \textbf{Type ID} & \textbf{Type} & \textbf{Description} & \textbf{Cost} \\ 
    \midrule
    \multirow{4}{*}{Node} & 0 & $v_c$ & Central node &  \\
    & 1 & $v_{cc}$ & Dummy node between central nodes & \\
    & 2 & $v_p$ & Proxy node, including $v_l, v_r$ &  \\
    & 3 & $v_{cp}$ & Dummy node between influx/efflux node and central node & \\
    \midrule
    \multirow{2}{*}{Edge} & 0 & $e_c$ & Edge between $v_c$, $v_{cc}$, a segment of central edges & $\alpha$ \\
    & 1 & $e_p$ & Edge concerning $v_p$ or $v_{cp}$, a segment of influence edges  & weight($e_p$) \\
    \midrule
    \multirow{4}{*}{Crossing} & 0 & & All crossing & \multirow{4}{*}{$\text{cost}(e_1) \cdot \text{cost}(e_2)$} \\
    & 1 & & Weighted crossing & \\
    & 2 & $(e_f, e_f)$ & Internal crossing & \\
    & 3 & $(e_p, e_p)$ & External crossing & \\
    \bottomrule
  \end{tabu}
\end{table}

During the layout of the GF graph, we adhere to the rules that influx/efflux nodes are always positioned on the left and right sides, arranged by year. In detail, we enhanced the Hierarchical Layout provided by GraphViz (DOT) to lay out the whole graph with influence edges and influx/efflux nodes rather than the central graph. As shown in Table \ref{tab:nodes_edges_crossings}, we considered fine-grained node types during the layout process. Due to the addition of dummy nodes in the layout process, node types include central nodes \( v_c \), influx/efflux nodes \( v_p \), dummy nodes \( v_{cc} \) between central nodes, and dummy nodes \( v_{cp} \) between central and influx/efflux nodes. Edges are divided into segments of central edges \( e_c \) and segments of influence edges \( e_p \). During the node ordering step, we ensure that influx nodes \( v_l \) and efflux nodes \( v_r \) are positioned at the extreme left and right of the layout, respectively, as highlighted in Algorithm \ref{alg:node_ordering}.

\begin{algorithm}[H]
\caption{Node Ordering with Constraint of Proxy Nodes}
\begin{algorithmic}[1]
\State \textbf{Input:} Graph $g$
\State \textbf{Output:} Best order $best$
\State order = \textsc{init\_order}(g)
\State best = order
\For{i = 0 to Max\_iterations}
    \State \textsc{wmedian}(order, i) \Comment{Calculate the median positions for each layer and sort nodes based on these positions, alternating the direction of each iteration}
    \State \textsc{transpose}(order) \Comment{Swap adjacent nodes to reduce edge crossings layer by layer}
    
    \For{r = 1 to Max\_rank}
        \State adjust\_order(order[r], $v_l$(r), $v_r$(r)) \Comment{Adjust order[r] to place $v_l$(r) at the start and $v_r$(r) at the end, shifting other nodes accordingly}
    \EndFor
    
    \If{\textsc{crossing}(order) $<$ \textsc{crossing}(best)}
        \State best = order
    \EndIf
\EndFor
\State \Return best
\end{algorithmic}
\label{alg:node_ordering}
\end{algorithm}

\subsection{Weighted Crossing}

The optimization metric is changed from edge crossing to weighted crossing. The weight of edges in the middle is set to \( \alpha \), and the weight of influence edges is proportional to edge width. In the crossing function, we use \( \text{cost}(e) \) instead of the default "1" as the edge weight. The weighted crossing between \( e_1 \) and \( e_2 \) is defined as \( \text{cost}(e_1) \cdot \text{cost}(e_2) \). We use total weighted crossing as the optimization objective, as highlighted in Algorithm \ref{alg:weighted_crossing}.

\begin{algorithm}[H]
\caption{Weighted Crossing Count Calculation}
\begin{algorithmic}[1]
\State \textbf{Input:} Order $order$
\State \textbf{Output:} crossing count $count$
\State count = 0
\State Count = \{\} \Comment{Count of connections to node $i$ in the next layer}
\For{r = 1 to Max\_rank}
    \For{v in order[r]}
        \For{e in \textsc{ND\_out}(v)}
            \For{k = order(\textsc{aghead}(e)) to max(keys(Count))}
                \State $count += Count[k] \times \text{cost}_\text{weighted}(e)$
            \EndFor
            \State Count[order(\textsc{aghead}(e))] += $\text{cost}_\text{weighted}(e)$
        \EndFor
    \EndFor
\EndFor
\State \Return count
\end{algorithmic}
\label{alg:weighted_crossing}
\end{algorithm}

To evaluate the optimal weight \( \alpha \) for the best layout, we used crossing counts as indicators, including internal crossing (between central edges) and external crossing (between influence edges). We tested four optimization objectives: all crossing, internal crossing, external crossing, and weighted crossing, defined as \eqref{eq:crossing}.

\begin{equation}
\begin{aligned}
    \text{cost}_\text{all}(e) &= 1 \\
    \text{cost}_\text{internal}(e) &= 
    \begin{cases} 
        1 & \text{if } e \in E(T_k) \\
        0 & \text{if } e \in E_I(T_k) 
    \end{cases} \\
    \text{cost}_\text{external}(e) &= 
    \begin{cases} 
        0 & \text{if } e \in E(T_k) \\
        1 & \text{if } e \in E_I(T_k) 
    \end{cases} \\
    \text{cost}_\text{weighted}(e) &= 
    \begin{cases} 
        \alpha & \text{if } e \in E(T_k) \\
        \text{weight}(e) & \text{if } e \in E_I(T_k) 
    \end{cases}
    \label{eq:crossing}
\end{aligned}
\end{equation}

\begin{figure}[tb]
  \centering 
  \includegraphics[width=\columnwidth]{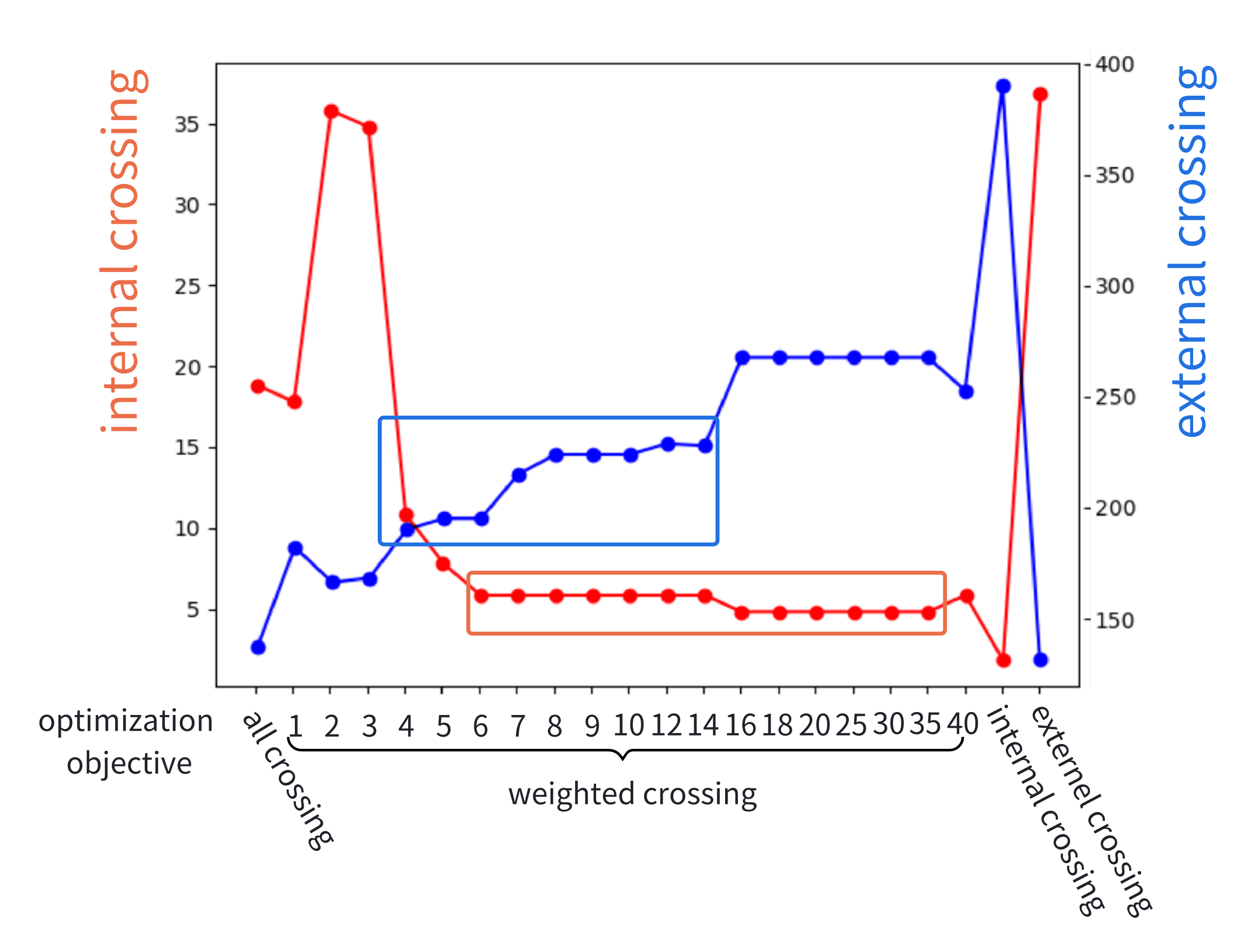}
  \caption{%
    Internal and external crossings under different optimization objectives, in an example graph of 50 nodes, are used for determining the optimal weight ($\alpha$) using the elbow method. 
  }
  \label{fig:indicator}
\end{figure}

We aimed to find an optimal objective where internal crossing was low (primary goal) while maintaining low external crossing. Using the elbow method, we determined the optimal \( \alpha \). For example, in a graph with 50 nodes, \rfig{indicator} shows internal and external crossings under different optimization objectives. Using "all crossing" as the baseline resulted in low external crossing but high internal crossing. "Internal crossing" had the lowest internal crossing but high external crossing, while "external crossing" had the opposite effect. Weighted crossing achieved a good balance; with \( \alpha \) over 6, internal crossing is relatively low, and further increasing \( \alpha \) provided no benefit while the external crossing arises extensively. Thus, we selected \( \alpha = 6 \) as the optimal weighted crossing for this example.

To generalize, we similarly selected \( \alpha \) incrementally, stopping when the reduction in internal crossing did not outweigh the increase in external crossing. This algorithm calculates an adaptive \( \alpha \), resulting in an enhanced topic GF graph with detailed influence information and minimal deviation from the original layout.

\subsection{Citation Influence Flow Map}

\begin{algorithm}[H]
\caption{Flow Map Adjustment Algorithm}
\begin{algorithmic}[1]
\State \textbf{Input:} Edge bundles $contextEdges$, Layout of Influence edges $layout$
\State \textbf{Output:} Adjusted flow map layout

\State $pointTree \gets$ buildPointTree($layout$)
\State $order \gets$ topologicalSort($pointTree$)

\For{each $p$ in $order$}
    \State $totalWidth \gets$ sum of widths of all paths at $p$
    \State $parentEdge \gets$ find parent edge from $pointTree$
    \State $parentEdge.width \gets totalWidth$
    \State $normal \gets$ computeNormal($parentEdge$)

    \State $sortedPaths \gets$ sort paths at $p$ by weighted angles ($0.5 \times angle1 + 0.3 \times angle2 + 0.2 \times angle3$)
    \State $point \gets$ initial position based on $p$ and $normal$

    \For{each path in $sortedPaths$}
        \State adjustStartPoint($path$, $point$, $normal$)
        \State update $point$ for next path
    \EndFor
\EndFor

\State \Return adjusted $layout$
\end{algorithmic}
\end{algorithm}

The flow map layout ensures that the edge widths increase progressively at intersection points, akin to a Sankey diagram. The primary goal is to prevent visual clutter and edge crossings for graphs with influence edges (\rfig{HierarchicalLayout}(b)), resulting in a clear and comprehensible visualization of complex network flows. To achieve the Influx/Efflux flow map, we first bundle the influence edges (\rfig{HierarchicalLayout}(c)), calculating their intersection points. Through modeling these intersection points, we adjust the layout by tweaking the positions, thicknesses, and order of the split edges, thereby achieving the effect of a Sankey graph (\rfig{HierarchicalLayout}(d)).

We handle edge bundling by merging adjacent dummy nodes into a single node. However, direct merging could confuse central edges with influence edges. Therefore, central edges and influence edges are treated separately. Not bundling central edges made the pattern more evident, while context information was less critical. Therefore, we achieved edge bundling by merging only influence edges (\( v_{cc} \) or \( v_{cp} \)) to simplify visual elements and highlight the topic GF graph.

To adjust the flow map layout based on edge bundling results, we follow several detailed steps. The edge bundling output consists of control points for Bézier curves, where the control points at both ends of these Bézier curves might coincide at the same point, indicating a branching point known as an intersection. To draw the flow map and create a Sankey diagram based on these edge bundling control points, we need to model these intersections through the following steps:

\begin{itemize}
    \item We construct an intersection tree from the pre-arranged curves.
    \item We perform a topological sort on the intersection tree to ensure that edge widths increase progressively from parent to child. This sort arranges the intersections in a hierarchical order, allowing for an organized flow of edges.
    \item Starting from the downstream flow (i.e., the end of the topological sort), we progressively adjust the curves upstream.
\end{itemize}

At each intersection, the total width of all curves is calculated by summing the widths of all paths. The normal direction at the intersection is then determined using the curve of the parent intersection. Curves at each child intersection are sorted using weighted angles, calculated based on the tangent lines formed by the curve's closest three points to the intersection, with the closest point having higher weights because these Bézier curves are influenced more by the points closest to the intersection. Finally, the starting points of curves at each child intersection are adjusted based on the normal direction and the sorted order.

The flow map effectively visualizes complex network interactions by combining edge bundling and layout adjustment. Progressive edge widths and clear separation of line flows are achieved, avoiding visual clutter and crossings. This approach realizes a control-point-based Sankey diagram, providing an efficient method for visualizing intricate networks and their interactions.